\documentclass[pdflatex,sn-mathphys-num]{sn-jnl}

\usepackage{graphicx}
\usepackage{multirow}
\usepackage{amsmath,amssymb,amsfonts}
\usepackage{amsthm}
\usepackage{mathrsfs}
\usepackage[title]{appendix}
\usepackage{xcolor}
\usepackage{textcomp}
\usepackage{manyfoot}
\usepackage{booktabs}
\usepackage{algorithm}
\usepackage{algorithmicx}
\usepackage{algpseudocode}
\usepackage{listings}
\usepackage{tikz}
\usetikzlibrary{positioning,arrows}
\usepackage{listings}
\usepackage{array}
\usepackage{caption}
\usepackage{float}
\usepackage{ragged2e}
\newcolumntype{P}[1]{>{\RaggedRight\arraybackslash}p{#1}}

\theoremstyle{thmstyleone}%
\theoremstyle{thmstyletwo}%

\theoremstyle{thmstylethree}%

\begin{document}

\title[Article Title]{A Thermodynamic Model for the Emergence of Natural Selection in Prebiotic Reaction Networks}


\author*[1]{\fnm{T. M.} \sur{Prosser}}

\abstract{The origin of life is often approached through the lens of replication, heredity, or molecular specificity. This paper proposes a thermodynamic framework in which the emergence of life is driven by the persistence of reaction pathways that align energetically with fluctuating environmental inputs. We define a reaction viability inequality based on energy input, release, resilience, and expenditure, which selects for persistent chemical configurations without invoking heredity or genetic encoding. We further incorporate entropic dynamics and spatial constraints into an augmented persistence function, showing that systems far from equilibrium can simultaneously increase global entropy while supporting localized chemical order. These refinements lead to the development of the \textbf{Thermodynamic Abiogenesis Likelihood Model (TALM)}, a probabilistic extension that estimates the likelihood of persistence-driven selection under diverse prebiotic and planetary scenarios. This framework redefines the conditions under which life-like organization may emerge and provides a testable, general theory for abiogenesis grounded in physical law.}


\maketitle

\section{Introduction}\label{sec1}

The origin of life remains one of the most challenging problems in science. While considerable progress has been made in identifying plausible prebiotic chemistries and environmental settings, a general physical principle that explains how lifelike organization can first emerge remains elusive. Existing models variously emphasize autocatalytic networks \cite{kauffman1993origins}, dissipative adaptation \cite{england2013statistical}, kinetic stability \cite{pross2012what}, and amphiphilic self-assembly into membranes and compartments \cite{tanford1980hydrophobic, deamer2011first}. Each of these approaches captures an aspect of the problem, but none provide an operational criterion for assessing the persistence of chemical systems in fluctuating prebiotic environments.

Recent theoretical work \cite{Kolchinsky2025} has examined the thermodynamic limits of Darwinian evolution in molecular replicators, deriving bounds that relate energy dissipation to differential reproductive success. While that analysis addresses systems in which heredity and competition are already established, the present study concerns a distinct and antecedent regime, one in which persistence itself functions as a selection filter prior to the emergence of replication. In this sense, the two approaches are complementary. Kolchinsky’s work formalizes thermodynamic constraints on post-replicative evolution, while the present model examines how selection-like\footnote{In this work the term selection-like refers solely to differential persistence arising from thermodynamic compatibility in fluctuating environments and does not imply adaptive fitness, heredity, or biological evolution.
} dynamics could arise spontaneously from energy-structure interactions in prebiotic chemistry.

Here we propose a theoretical framework in which selection-like behavior emerges without heredity or replication, driven instead by thermodynamic constraints on energy input, use, storage, and dissipation. This model formalizes a statistical pathway to abiogenesis rooted in the persistence of reaction networks whose viability depends on environmental energy availability, internal energy dynamics, and structural resilience. This formulation treats natural selection as a consequence of energy-compatible persistence. It links entropy production, spatial organization, and reaction viability within a physically grounded, molecule-agnostic framework. In this view, selection-like behavior can emerge in any system where energy regulation supports sustained reaction intervals.

This work develops a thermodynamic model in which the origin of life is understood as the emergence of persistent chemical structure within energy-modulated environments. The model assumes no initial heredity, replication, templating, enzymatic catalysis, or specific molecular identity. Grounded in first principles and agnostic to molecular identity, the framework describes how abiogenesis may arise from prebiotic reaction chains shaped by entropy flux and diffusion constraints.

We extend this framework into a probabilistic formulation, the Thermodynamic Abiogenesis Likelihood Model (TALM), which estimates the likelihood of persistence-driven selection under diverse planetary conditions. TALM characterizes the emergence of persistence-based selection in localized chemical environments and can be extended to diverse planetary settings. The remainder of this paper develops the model’s thermodynamic foundation, introduces its statistical structure, and examines its implications for prebiotic chemical organization.

As an initial test of the model’s predictions, we analyze a comparative system involving amphiphilic molecules of varying chain length. This system isolates entropy and structural persistence as primary variables and allows for experimental validation of the core mechanisms under well-defined aqueous conditions.
\newpage
\section{Terminology and Variable Definitions}

To aid clarity and interdisciplinary accessibility, the following table summarizes key variables and terms used in the model\footnote{Molecular-scale values are expressed in J per molecule; values in the main text use J·mol–1 for clarity, convertible via Avogadro’s number.}:

\begin{table}[h!]
\centering
\caption{Symbols, units, and descriptions used in the framework.}
\begin{tabular}{|p{2cm}|p{2cm}|p{9cm}|}
\toprule
\textbf{Symbol} & \textbf{Units} & \textbf{Description} \\
\midrule
$k$ & dimensionless & Number of reactions in the reaction chain. \\
\midrule
$z(t)$ & J·mol$^{-1}$ & Time-varying environmental energy input available to the system. \\
\midrule
$y(t)$ & J·mol$^{-1}$ & Residual energy at time $t$; positive values indicate persistence. \\
\midrule
$x_i$ & J·mol$^{-1}$ & Energy required to perform reaction $i$ (activation or maintenance cost). \\
\midrule
$r_i$ & J·mol$^{-1}$ & Energy released as usable output from reaction $i$. \\
\midrule
$s_i$ & J·mol$^{-1}$ & Potential stored energy in reaction products that can contribute to $S(t)$. \\
\midrule
$S(t)$ & J·mol$^{-1}$ & Stored energy within the system; evolves according to $\frac{dS}{dt} = \sum_i s_i - U(S,t)$. \\
\midrule
$U(S,t)$ & J·mol$^{-1}$·s$^{-1}$ & Release or utilisation rate of stored energy as a function of $S(t)$ and time. \\
\midrule
$R_n$ & J·mol$^{-1}$ &
Resilience contribution, defined as $R_n = \lambda_R \mathcal{R}(G)$, 
where $\mathcal{R}(G)\!\in\![0,1]$ is a normalized redundancy factor from network topology 
and $\lambda_R$ is an energy scale (typically $RT$ per mole). \\
\midrule
$\Delta S_{\text{net}}$ & J·K$^{-1}$·mol$^{-1}$ & Net entropy change ($\Delta S_{\text{env}} - \Delta S_{\text{local}}$) between system and environment. \\
\midrule
$T\Delta S_{\text{net}}$ & J·mol$^{-1}$ & Entropy term expressed in energy units at temperature $T$. \\
\midrule
$D(x,t)$ & m$^{2}$·s$^{-1}$ & Diffusion coefficient describing spatial dispersion or confinement. \\
\midrule
$\alpha$ & J·mol$^{-1}$·K & Scaling coefficient for the entropy-related penalty contribution. \\
\midrule
$\beta$ & J·mol$^{-1}$·s·m$^{-2}$ & Scaling coefficient for the diffusion-related penalty contribution. \\
\midrule
$\Phi(t)$ & J·mol$^{-1}$ & Entropic-diffusive penalty term $\alpha_S T \Delta S_{\text{local}}(t) + \beta_D \dfrac{R T}{D_0} D(x,t)$. \\
\midrule
$y'(t)$ & J·mol$^{-1}$ & Augmented persistence function $y'(t) = y(t) + R_n - \Phi(t)$. \\
\midrule
$P_{\text{selective}}$ & dimensionless & Probability of persistence across $N$ reaction permutations: $P_{\text{selective}} = 1 - (1 - P(y'(t)\ge 0))^N$. \\
\botrule
\end{tabular}
\end{table}

\noindent
\newline

\newpage

\section{The Abiogenesis Persistence Framework}

\subsection{Assumptions and Principles}
This model is constructed on a set of thermodynamic principles and underlying assumptions about how energy is used, transferred, and stored in prebiotic chemical systems. These assumptions are not intended to capture the full complexity of biochemistry, but rather to provide a minimalist framework capable of illustrating how selection-like dynamics may arise from energy-driven processes alone.

\subsection{Defining Persistence}

Persistence, as used in this model, is not introduced as a fundamental thermodynamic state variable. Instead, it represents an emergent property of a driven chemical system: the ability to maintain non-zero reaction fluxes, concentration gradients, or structural organisation over time in the presence of dynamic environmental input \cite{Prigogine1972}. Such persistence reflects the system’s sustained capacity to export entropy to its surroundings and to maintain a favourable balance between energy uptake, dissipation, and loss \cite{schroedinger1944life}. Related concepts such as dynamic kinetic stability similarly treat extended molecular or network lifetimes as consequences of energy flux and reaction kinetics rather than as primitive thermodynamic variables \cite{pross2012what}. Experimental studies on wet–dry cycling and protocell formation likewise show that structural persistence in prebiotic contexts depends critically on continuous environmental driving \cite{song2012wetdry}. In this framework, persistence serves as a practical, derived descriptor for assessing whether prebiotic reaction systems remain viable over time.

\subsection{Energy Availability and Use}
We assume that reactions in prebiotic systems are governed by basic thermodynamic constraints:
\begin{itemize}
    \item Every chemical reaction has an associated energy threshold (activation energy) that must be met for the reaction to proceed.
    \item A system only uses as much energy as is required to complete a reaction step. If more energy is available than needed, the surplus may be either released (e.g., as heat or dissipated energy), or stored in the reaction products (e.g., as chemical potential energy).
\end{itemize}

\subsection{Dynamic Environmental Input}
We introduce a time-dependent energy function, $z(t)$, representing the usable energy made available from the environment. This can fluctuate due to factors such as sunlight cycles, geothermal activity, or chemical gradients. This treatment reflects empirical work showing that surface thermal cycling, particularly in hydrothermal field environments, can generate temporally structured energy availability conducive to prebiotic polymerization \cite{deamer2011first} \cite{ross2019drywet}.  The persistence of a reaction chain is therefore not only a function of its internal energy efficiency but also its ability to maintain viability under changing energy input.
\subsection{System Definition}
Let the system consist of a series of $k$ reactions, each requiring a defined amount of energy to proceed and potentially releasing or storing energy in the process. We define the following terms:
\begin{itemize}
  \item $k$: The number of reactions in the chain
  \item $z(t)$: The total usable energy available to the system from the environment at time $t$.
  \item $x_i$: The energy required to perform reaction $i$.
  \item $r_i$: The energy released during reaction $i$.
  \item $s_i$: The potential stored energy in reaction products.
  \item $S(t)$: Stored free energy within the system.
  \item $y(t)$: Residual energy at time $t$.
\end{itemize}

\subsubsection{Reaction Chains as Systems}
Rather than modeling isolated reactions, we conceptualize reaction chains, sequences of reactions where the output or byproducts of one reaction may fuel or enable subsequent ones. These chains can be:
\begin{itemize}
    \item Autocatalytic (self-amplifying)
    \item Branching or linear
    \item Open to energy inflow (from the environment via $z(t)$)
\end{itemize}
A chain will continue as long as the net energy available (initial input plus internal releases) meets the sum of energy requirements across the chain.
\subsubsection{The Reaction Viability Inequality}
For a system to persist at time $t$, the net energy budget must meet or exceed the energy costs required to sustain the chain of reactions:
\begin{equation}
y(t) = z(t) + S(t) + \sum_{i=1}^{k} r_i - \sum_{i=1}^{k} x_i
\end{equation}

Persistence is determined by:
\begin{equation}
y(t) \geq 0 \Rightarrow \text{System persists}
\end{equation}

If this condition is not met:
\begin{equation}
y(t) < 0 \Rightarrow \text{System collapses}
\end{equation}

Here, $k$ is the number of reactions in the chain. This inequality acts as a filter: only those reaction chains whose internal energy efficiencies align with external energy fluctuations will persist. This creates an emergent selection pressure favoring systems with adaptive, energy-efficient structures.
\subsubsection{Energy Storage Dynamics and Future Use (S(t))}

In open prebiotic systems, a fraction of the energy released during reactions can be retained in chemical or structural form rather than dissipated. This potential stored energy ($s_i$) is not immediately usable unless another reaction releases it.  Thus, stored energy acts as a possible buffer or reserve, contributing to a system's future viability but not necessarily to its current energy budget.  To represent this explicitly, a storage variable $S(t)$ is introduced to track the time-dependent accumulation and release of stored energy within the system. The rate of change of stored energy is expressed as

\[
\frac{dS}{dt} = \sum_i s_i - U(S,t)
\]

where $\sum_i s_i$ is the total energy stored in reaction products and $U(S,t)$ is the energy released or utilized from storage as a function of the current state $S(t)$ and environmental conditions. The total available energy at time $t$ therefore becomes

\[
y(t) = z(t) + \sum_i r_i - \sum_i x_i + S(t)
\]

which ensures that all sources and sinks of energy are explicitly accounted for over a reaction cycle. This formulation preserves conservation of energy across fluctuating input conditions, allowing the persistence inequality $y'(t) \ge 0$ to be interpreted as a full-cycle energy balance in which stored energy may buffer temporary deficits in environmental supply.

\begin{figure}[htbp]
\centering
  \includegraphics[width=1\textwidth]{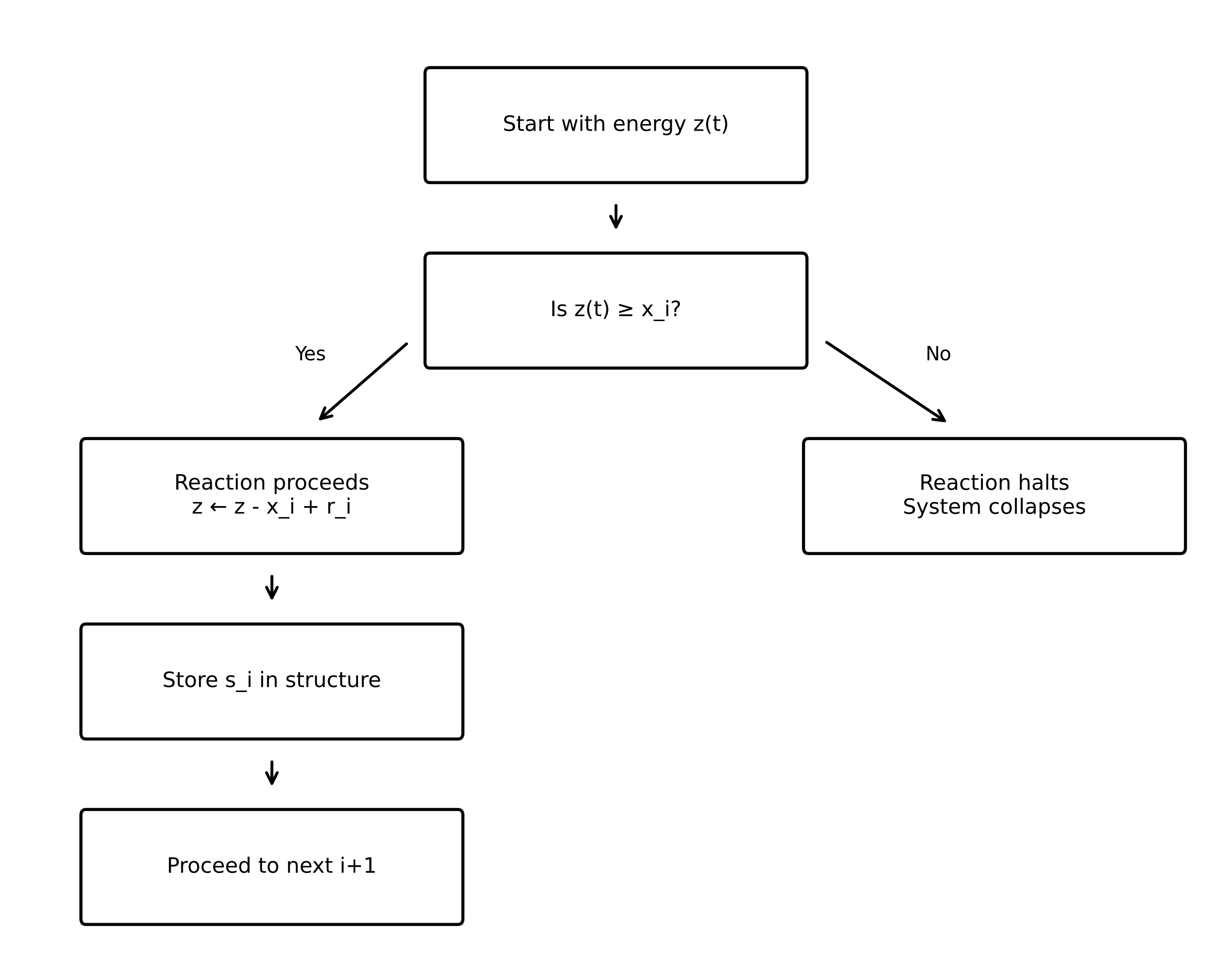}
  \caption{Figure~\ref{fig:flowchart} illustrates the logic of the reaction chain model under fluctuating environmental energy.  The flowchart summarizes the operational logic of the reaction-chain persistence model under fluctuating environmental energy.}
  \label{fig:flowchart}
\end{figure}

\subsection{Modulators of Persistence}

The persistence inequality describes the basic energetic requirement for the continuation 
or collapse of a reaction sequence. In realistic prebiotic settings, however, this energy 
balance is influenced by additional physical factors that alter either the effective 
energy available to a system or the losses it incurs. These modulators include dimensional 
constraints on the quantities entering the inequality, the dynamics of energy storage and 
release, the entropic costs associated with local ordering, spatial and diffusive effects, 
and the resilience provided by network redundancy. Together, these terms extend the core 
inequality into a form suitable for analyzing prebiotic systems subject to fluctuating, 
noisy, or spatially structured environments.

\subsubsection{Resilience Contribution (\texorpdfstring{$R_n$}))}

The term $R_n$ represents the contribution of network redundancy or structural buffering 
to the overall viability of a reaction sequence. It accounts for the fact that certain 
reaction networks can sustain temporary local failures, dispersal events, or energetic 
shortfalls if alternative pathways or parallel processes remain available. To ensure 
dimensional coherence with the persistence inequality, $R_n$ is defined as
\[
R_n = \lambda_R \,\mathcal{R}(G),
\]
where $\lambda_R$ is an energy scale (taken here as $RT$ per mole at a reference 
temperature) and $\mathcal{R}(G)\!\in\![0,1]$ is a normalized measure of network 
redundancy derived from the topology $G$. Higher values of $\mathcal{R}(G)$ correspond 
to reaction networks in which multiple reaction sequences or parallel pathways provide 
buffering against the loss of individual steps. In this way, $R_n$ captures the 
thermodynamic advantage conferred by structural resilience without altering the basic 
form of the persistence inequality.

\subsubsection{Entropy, Diffusion, and Spatial Organization}

Entropy production in open thermodynamic systems enables exploration of chemical configuration space, while spatial structure constrains how this exploration unfolds \cite{schneider2005into, prigogine1977self}. Local decreases in entropy associated with ordering or assembly must be compensated by entropy export to the surroundings. We therefore define the net entropy balance as:
\[
\Delta S_{\text{net}} = \Delta S_{\text{env}} - \Delta S_{\text{local}}.
\]

Here, we take $\Delta S_{\text{env}} \ge 0$ to denote entropy exported to the environment, and $\Delta S_{\text{local}} \ge 0$ to denote the magnitude of entropy reduction associated with local ordering and thus the feasibility condition
$\Delta S_{\text{net}} = \Delta S_{\text{env}} - \Delta S_{\text{local}} > 0$
ensures compatibility with the second law - systems that decrease local entropy without adequate environmental compensation collapse even if energy thresholds are otherwise met.

Diffusion $D$ further modulates viability. High diffusivity disperses reactants and products, increasing the likelihood of network collapse, whereas confinement in vesicles, pores, or mineral surfaces lowers effective diffusivity and promotes retention. These qualitative effects motivate the introduction of an entropic-diffusive penalty term in the augmented viability function, developed in the next subsection.

Entropy balance and diffusion can be incorporated without altering the structure of the core inequality. We represent the effects through an entropic–diffusive penalty term $\Phi(t)$, which enters the augmented viability function
\[
y'(t) = y(t) + R_n - \Phi(t).
\]

The penalty term $\Phi(t)$ combines the energetic cost of maintaining local order with the
losses associated with molecular dispersal. A unit–consistent representation is
\[
\Phi(t) = \alpha_S T\Delta S_{\mathrm{local}}
        + \beta_D \frac{RT}{D_0} D(x,t),
\]
where $\alpha_S$ and $\beta_D$ are dimensionless scaling factors,
$T\Delta S_{\mathrm{local}}$ represents the energetic cost of local entropy
reduction, and $(RT/D_0) D(x,t)$ provides a diffusion‐scaled energetic loss
relative to a reference diffusivity $D_0 = 10^{-9}\,\mathrm{m^2\,s^{-1}}$.
This ensures that $\Phi(t)$ retains units of energy per mole and remains directly
comparable with $y(t)$ and $R_n$.  More detailed models could replace $\Phi(t)$ with system-specific expressions without altering the role of the viability boundary.  In this way, the augmented formulation embeds entropy export, spatial confinement, and structural resilience directly into the viability boundary, extending the thermodynamic analysis to realistic prebiotic environments while preserving the simplicity of the core inequality.

\subsection{Thermodynamic Consistency and Gibbs Free Energy Formulation}

The persistence inequality is fundamentally an energy budget: a reaction network persists when the free energy supplied by the environment, together with any recoverable stored or structural energy, is sufficient to offset the free-energy costs of reaction steps and the losses associated with entropy reduction and diffusion. This condition can be expressed more compactly and transparently using Gibbs free energy, which naturally incorporates both enthalpic and entropic contributions to reaction viability.

For a discrete reaction cycle $n$, we define the net free-energy change of the system as
\[
\Delta G_{\mathrm{net},n} 
= \Delta G_{\mathrm{in},n}
+ \sum_{i=1}^k \Delta G_{x,i}
+ \sum_{i=1}^k \Delta G_{r,i}
+ \Delta G_{\Phi,n}
- \Delta G_{R,n}.
\]

Each term represents a free-energy contribution per cycle, with the following sign conventions:

\begin{itemize}
    \item $\Delta G_{\mathrm{in},n}$ is the effective free energy made available to the system from the environment. For a favourable inflow, $\Delta G_{\mathrm{in},n} < 0$.
    \item $\Delta G_{x,i} > 0$ is the free-energy cost associated with initiating or maintaining reaction $i$.
    \item $\Delta G_{r,i}$ represents the free-energy change associated with reaction releases. When a reaction releases usable free energy, $\Delta G_{r,i} < 0$.
    \item $\Delta G_{\Phi,n} > 0$ is the free-energy penalty associated with local entropy reduction and molecular diffusion, derived from the entropic–diffusive term $\Phi(t)$.
    \item $\Delta G_{R,n} > 0$ is the recoverable free-energy contribution arising from structural resilience or redundancy in the network. This term is subtracted because it offsets net free-energy costs.
\end{itemize}

Persistence requires that the net free-energy change of the system be non-positive:
\[
\boxed{\Delta G_{\mathrm{net},n} \le 0.}
\]

This expresses the requirement that the system, considered over one complete cycle of environmental driving, either dissipates free energy or at least does not accumulate a positive free-energy deficit. Reaction networks with $\Delta G_{\mathrm{net},n} > 0$ cannot maintain the fluxes and entropy production required for persistence and therefore collapse.

Mapping between the free-energy formulation and the original energy-budget notation is straightforward:
\[
z_n \;\leftrightarrow\; -\,\Delta G_{\mathrm{in},n},\qquad
x_i \;\leftrightarrow\; \Delta G_{x,i},\qquad
r_i \;\leftrightarrow\; -\,\Delta G_{r,i},\qquad
R_n \;\leftrightarrow\; \Delta G_{R,n},\qquad
\Phi_n \;\leftrightarrow\; \Delta G_{\Phi,n}.
\]

Thus the viability condition $y'_n \ge 0$ is directly equivalent to the thermodynamically grounded condition $\Delta G_{\mathrm{net},n} \le 0$. The two formulations differ only in representation: the former uses explicit energy accounting, while the latter embeds the same quantities within the standard Gibbs framework for open, driven chemical systems.

\subsubsection{Open-System Energy Balance}

The persistence inequality can be viewed as a simplified expression of the first law of
thermodynamics for an open system in which energy flows continuously between the
environment and the reaction network. At any time $t$, the total energy change of the
system can be written in the coarse-grained form
\[
\frac{dU_{\text{sys}}}{dt}
= \dot{Q}_{\text{in}} - \dot{Q}_{\text{out}}
+ \dot{W}_{\text{in}} - \dot{W}_{\text{out}}
+ \frac{dS}{dt},
\]
where $\dot{Q}$ represents heat exchange with the environment, $\dot{W}$ represents
non-thermal work terms associated with reaction processes, and $\frac{dS}{dt}$ is the
net rate of change in stored free energy $S(t)$ defined earlier in the framework. Here
$d$ denotes the differential operator, so $\frac{dS}{dt}$ is the continuous-time analogue
of the discrete update of $S(t)$.

Persistence requires that the sum of incoming energy and recoverable stored energy
meets or exceeds the outgoing dissipative terms. Expressed at the scale of reaction
events, this condition reduces to the viability inequality
\[
y'(t) = z(t) + \sum_i r_i - \sum_i x_i + S(t) + R_n - \Phi(t) \ge 0,
\]
linking macroscopic energy balance to the microscopic dynamics of reaction viability.
This relationship ensures that the framework remains consistent with thermodynamic
constraints while retaining a form simple enough to apply across diverse prebiotic
environments.

\subsubsection{Dimensional Consistency and Energy Balance}

To maintain unit coherence within the persistence inequality, all variables are expressed in standard thermodynamic dimensions. The energy terms $x_i$, $r_i$, and $s_i$ represent energy per reaction event, measured in joules per mole (J·mol$^{-1}$). The environmental input $z(t)$ and residual energy $y(t)$ share the same units. Entropic terms such as $T\Delta S_{\text{net}}$ are expressed in joules per mole through multiplication of temperature (K) by entropy (J·K$^{-1}$·mol$^{-1}$). The diffusion factor $D(x,t)$ has dimensions of m$^{2}$·s$^{-1}$, and the entropic-diffusive penalty $\Phi(t)$ retains units of energy per mole to remain directly comparable with $y(t)$. 

A simple consistency check can be demonstrated by substituting representative values: if a reaction requires $x_i = 5\times10^{-20}$ J·molecule$^{-1}$, releases $r_i = 2\times10^{-20}$ J·molecule$^{-1}$, and the system experiences an environmental input $z(t) = 9\times10^{-20}$ J·molecule$^{-1}$, then the residual energy is $y(t) = z(t) + \sum r_i - \sum x_i = 6\times10^{-20}$ J·molecule$^{-1}$. The units of all terms resolve to energy, confirming dimensional closure of the reaction viability inequality. This ensures that the model remains physically grounded and that persistence, expressed as $y'(t) \ge 0$, is evaluated on an absolute energy basis consistent with open-system thermodynamic accounting.

\subsection{Relation to Non-Equilibrium Thermodynamic Formalisms}

The persistence framework can be situated within the broader developments
of non-equilibrium thermodynamics, from classical entropy production
theory to modern stochastic approaches. In the canonical formulation of
Prigogine and Nicolis, open systems far from equilibrium are governed by
generalized thermodynamic forces $X_i$ (e.g., chemical potential gradients)
and fluxes $J_i$ (reaction rates, diffusion currents), with entropy
production given by

\begin{equation}
\sigma = \sum_i J_i X_i \geq 0,
\end{equation}

\cite{Prigogine1972}. Dissipative structures arise when thermodynamic fluxes and forces organize to maintain $\sigma > 0$. Under these conditions, local decreases in entropy are possible, provided they are compensated by sufficient entropy export to the environment. In this tradition, our
condition $\Delta S_{\text{net}} > 0$ is directly compatible: persistence
requires that local ordering is offset by sufficient entropy release to the
environment. Collapse occurs when this balance fails, leading to vanishing
fluxes and extinction of the system.

In this setting, the viability condition $y'(t) \ge 0$ can be interpreted as the coarse-grained requirement that a reaction system maintains non-vanishing thermodynamic fluxes, and thus positive entropy production, over the relevant time interval. When $y'(t) < 0$, the system’s capacity to sustain the flux–force pairs $(J_i, X_i)$ diminishes, driving $\sigma \to 0$ as reaction rates collapse and the network loses the ability to export entropy. In this sense, the persistence inequality translates the entropy-production criterion $\sigma > 0$ for open, driven systems into a local survival condition for specific prebiotic reaction networks.

Recent developments in stochastic thermodynamics extend this picture to finite systems and fluctuating environments. These approaches quantify how entropy production behaves under stochastic forcing and how transient departures from the second law are statistically constrained. The Jarzynski equality\cite{jarzynski1997} and Crooks fluctuation theorem\cite{crooks1999} quantify the probability of transient reductions in entropy. Such events are possible, but they are exponentially suppressed, ensuring that the second law holds overwhelmingly on average.

Seifert’s review \cite{seifert2012} unifies these results, establishing that
entropy production remains the organizing constraint at Mesoscopic scales.
Within this context, the persistence inequality provides a coarse-grained
survival criterion: reaction networks persist only if $y'(t) \geq 0$ holds
with high probability across stochastic trajectories. This situates the
persistence filter as an operationalization of fluctuation theorems in the
specific domain of prebiotic reaction chains.

Esposito and Van den Broeck \cite{esposito2010} and Deffner \& Lutz
\cite{deffner2011}, further developed non-equilibrium frameworks for driven systems. Their work emphasizes fluctuation-dissipation relations and establishes efficiency bounds for systems undergoing external driving. Our framework differs in focus: rather than maximizing dissipation
or efficiency, it identifies a viability boundary where persistence either
succeeds or fails. This makes persistence not a variational outcome but a
filter imposed by environmental compatibility.

Finally, thermodynamic accounts of abiogenesis such as Michaelian’s dissipation theory \cite{michaelian2011,michaelian2023} propose that life arose as a mechanism for maximizing photon dissipation. This view highlights dissipation as central, but it focuses on a specific energy channel, unlike the more general energy-structure compatibility emphasized by the persistence inequality.  Dissipation is necessary but not sufficient: persistence requires meeting the viability criterion in dynamic environments.

Taken together, these connections show that the persistence framework is grounded in non-equilibrium thermodynamics. It condenses broad principles of dissipation, entropy production, and fluctuation constraints into a falsifiable, system-specific viability inequality.

\section{Building a Simulation}

\subsection{Simulation Logic}

The model is implemented numerically by evaluating the persistence inequality across a sequence of
time-steps. Each time-step represents an interval in which the system receives environmental free
energy, incurs internal energetic costs, and undergoes changes in stored energy, resilience, and
entropy-related dissipation.

At each interval, the free-energy balance is computed using the unified viability function,
\[
y(t) = z(t) + S(t) + \sum_i r_i - \sum_i x_i,
\]
and its thermodynamically refined form,
\[
y'(t) = y(t) + R(t) - \Phi(t).
\]

Environmental free energy $z(t)$ is drawn from a stochastic distribution representing variable
inputs such as temperature fluctuations, photochemical variability, or intermittent chemical
gradients. The quantities $x_i$ and $r_i$ are set by the reaction network and represent the energetic
costs and stabilising contributions specific to that configuration.

The entropic–diffusive penalty and structural resilience are evaluated directly:
\[
\Phi(t) = \alpha_S\,T\,\Delta S_{\text{local}}(t)
        + \beta_D\,\frac{RT}{D_0}\,D(t),
\qquad
R(t) = \lambda_R\,R(G(t)),
\]
ensuring consistency with the free-energy interpretation established in the theoretical framework.

Stored free energy evolves according to the discrete-time update,
\[
S(t+\Delta t) = S(t) + \sum_i s_i(t) - \gamma S(t),
\]
which is the time-stepped analogue of the continuous relation
\[
\frac{dS}{dt} = \sum_i s_i(t) - \gamma S(t).
\]
The term $\gamma S(t)$ captures generic losses such as leakage, instability, or uncaptured work.

Persistence occurs at a time-step whenever
\[
y'(t) \ge 0,
\]
which is equivalent to a non-positive net Gibbs free-energy change over that interval.
The proportion of intervals satisfying this condition across $N$ time-steps provides an estimate of
the persistence frequency. These values are used to construct the free-energy landscapes presented in
the results, illustrating how environmental variability, energetic costs, structural retention, and
entropy production jointly determine the likelihood of sustained reaction progression.

A computational model can be constructed to evaluate the persistence of reaction chains under fluctuating environmental energy input \( z(t) \), internal energy transformations, and thermodynamic constraints.

At each time step, the simulation proceeds as follows:
\begin{enumerate}
  \item Update \( z(t) \), the environmental energy input, based on a predefined or empirical function (e.g., sinusoidal, stochastic, or data-driven).
  \item For each reaction chain:
  \begin{itemize}
    \item Attempt reactions sequentially.
    \item If the energy requirement \( x_i \) is met, subtract it from the available energy, add the released energy \( r_i \), and store surplus energy \( s_i \).
    \item If the energy requirement is not met, the chain halts or collapses.
  \end{itemize}
  \item Chains that persist may optionally replicate, mutate, or recombine.
  \item The process repeats for the defined simulation window.
\end{enumerate}

Two simulation models are proposed here in order to show the core principle behind the persistence model, and an augmented simulation designed to show how iterations of the model can incorporate entropic and spatial factors. Further iterations may integrate a number of other factors and compartmentalization effects directly into chain viability evaluations.

\subsection{Parameter Ranges and Boundary Conditions}

To apply the persistence model in realistic simulations, we constrain the variables used in the augmented persistence inequality:

\[
y'(t) = y(t) + R_n - \Phi(t), \quad \text{where} \quad \Phi(t) = \alpha_S T\,\Delta S_{\text{local}}(t) + \beta_D \frac{R T}{D_0} D(x,t)
\]

Based on prebiotic and physical estimates, the following parameter ranges are proposed:

\begin{itemize}
  \item \textbf{Environmental Energy Input \(z(t)\)}:  
  Typical fluctuations range from \(10^{-21}\) to \(10^{-18}\,\mathrm{J\;molecule^{-1}}\) per event, reflecting contributions from UV irradiation, geothermal cycling, or chemical redox processes \cite{schroedinger1944life, morowitz1968energy}.  
  In the model, these values correspond directly to sampled realizations of \(z(t)\).

  \item \textbf{Reaction Energy Requirements \(x_i\)}:  
  Energetic costs for individual reaction steps are typically \(10^{-20}\) to \(10^{-18}\,\mathrm{J\;molecule^{-1}}\), representing bond formation, activation barriers, or structural maintenance.

  \item \textbf{Reaction Energy Releases \(r_i\)}:  
  Exergonic steps return energy to the system in the same units as \(x_i\) and \(z(t)\). Typical magnitudes fall within \(10^{-21}\) to \(10^{-19}\,\mathrm{J\;molecule^{-1}}\), depending on reaction class.

  \item \textbf{Entropy Balance \(\Delta S_{\mathrm{net}}\)}:  
  Net entropy export is positive when environmental dissipation  
  \(\Delta S_{\mathrm{env}} \sim 10^{-21} \text{ to } 10^{-20}\,\mathrm{J\,K^{-1}}\)  
  exceeds local ordering  
  \(\Delta S_{\mathrm{local}} \sim 10^{-22}\,\mathrm{J\,K^{-1}}\).  
  The viability penalty uses \(\Delta S_{\mathrm{net}}^{-1}\) through the function \(\Phi(t)\).

  \item \textbf{Diffusion Modifier \(D(x,t)\)}:  
  Effective diffusivity is environment dependent. Confined systems such as vesicles or mineral pores typically exhibit  
  \(D \sim 10^{-14}\) to \(10^{-12}\,\mathrm{m^{2}\,s^{-1}}\),  
  whereas bulk aqueous diffusion is approximately  
  \(D \sim 10^{-9}\,\mathrm{m^{2}\,s^{-1}}\).  
  Higher diffusivity increases \(\Phi(t)\).

  \item \textbf{Resilience Term \(R_n\)}:  
  A dimensionless parameter representing structural redundancy, buffering, or pathway multiplicity. Values range from 0 (no resilience) to 1 (high redundancy). Autocatalytic or feedback-rich networks may exhibit \(R_n > 0.5\).

  \item \textbf{Temporal Window \(t\)}:  
  Simulations commonly explore \(t\) from \(10\) to \(10^{5}\) seconds (seconds to days), consistent with prebiotic cycling intervals such as wet-dry phases, diurnal heating, or geothermal oscillations.
\end{itemize}

Boundary conditions may include:
\begin{itemize}
  \item Fixed or variable environmental energy sources
  \item Randomized initial reactant sets or network topologies
  \item Compartmental geometry or surface structure modifiers
\end{itemize}

Simulation outcomes can be interpreted in terms of the probability \( P(y'(t) \geq 0) \), and aggregated across permutations \( N \) to derive \( P_{\text{selective}} \) for differing alternative scenarios.

\subsubsection{Simulation defaults and reproducibility.}
Unless otherwise noted, simulations employ 
$\bar{z} \in [10^{-21},10^{-19}]$\,J\,molecule$^{-1}$, 
$\sigma_z = 0.2\,\bar{z}$, 
$D \in \{10^{-14},10^{-12},10^{-9}\}$\,m$^2$s$^{-1}$, 
$\gamma \in \{10^{-6},10^{-4}\}$\,s$^{-1}$, 
$\alpha_S = \beta_D = 1$, 
and $R_n = RT\,\mathcal{R}(G)$ at $T=330$\,K. 
Each parameter pair is evaluated using $N_{\mathrm{trials}}=10^3$ 
across $N_{\mathrm{cycles}}=10^3$ iterations. 
The resulting phase boundaries remain stable under $\pm 20\%$ variation, 
indicating robust qualitative behavior.

\subsection{Baseline Simulation (No Entropic or Spatial Constraints)}
The persistence inequality was implemented as a stochastic simulation to evaluate the fraction of viable reaction chains under varying environmental inputs. Each simulation represents an open system experiencing discrete energy intervals, during which the environmental input $z(t)$ fluctuates as a random variable drawn from a normal distribution with mean $\bar{z}$ and standard deviation $\sigma_z$. Stored energy $S(t)$ accumulates and dissipates across intervals according to $\frac{dS}{dt} = \sum s_i - \gamma S(t)$. The algorithmic form is given below.

\begin{algorithm}[H]
\caption{Baseline Persistence Simulation (No Entropy or Diffusion)}
\begin{algorithmic}[1]

\State $S \gets 0$    \Comment{Stored free energy}
\State $\mathcal{P} \gets \emptyset$    \Comment{Persistence record}

\For{each time-step $t$}

    \State $z \gets$ SampleEnvironmentalEnergy($t$)

    \For{each reaction chain $\mathcal{C}$}

        \State $E \gets z + S$     \Comment{Initial available free energy}
        \State $\mathcal{C}.\text{viable} \gets \text{True}$

        \For{each reaction $i$ in $\mathcal{C}$}

            \State $(x_i, r_i, s_i) \gets$ ReactionParameters($i$)

            \If{$E < x_i$}
                \State $\mathcal{C}.\text{viable} \gets \text{False}$
                \State \textbf{break}
            \EndIf

            \State $E \gets E - x_i + r_i$
            \State $S \gets S + s_i$

        \EndFor

        \State $\mathcal{P} \gets \mathcal{P} \cup \{(t, \mathcal{C}, \mathcal{C}.\text{viable})\}$

    \EndFor

    \State $S \gets S - \gamma S$   \Comment{Loss of stored free energy}

\EndFor

\State \Return $\mathcal{P}$

\end{algorithmic}
\end{algorithm}

\begin{figure}[h!]
\centering
\includegraphics[width=0.8\textwidth]{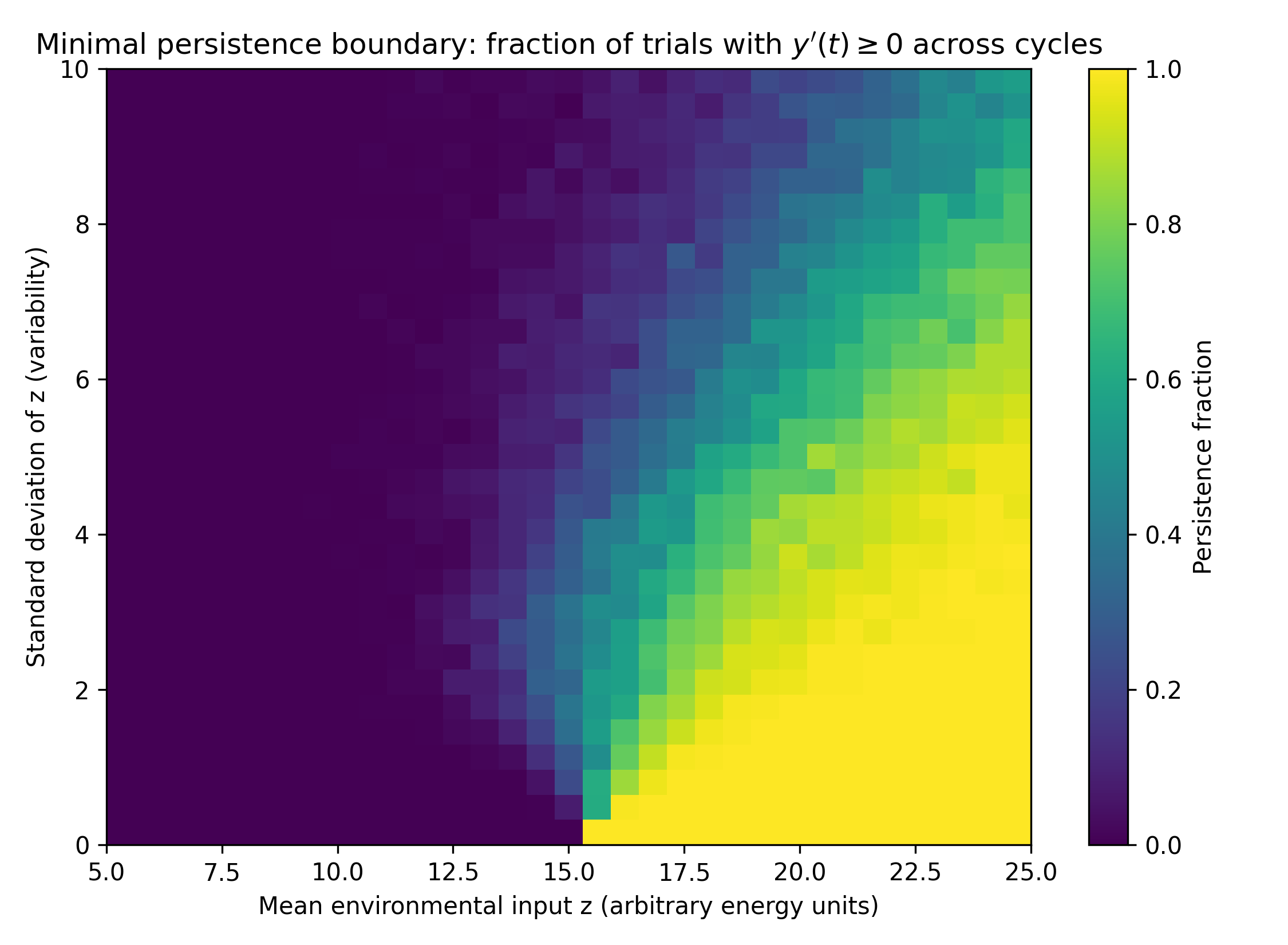}
\caption{Phase boundary between persistence and collapse predicted by the framework. The heatmap shows the fraction of simulated trials in which $y'(t)\geq0$ across multiple environmental intervals, plotted against the mean environmental energy input $\bar{z}$ and its variability $\sigma_z$. Higher $\bar{z}$ and lower variability favour persistence, demonstrating a clear transition between viable and non-viable regimes.}
\label{fig:persistenceboundary}
\end{figure}
\noindent
The baseline simulation captures persistence dynamics driven solely by variable energy input and storage.

\subsection{Augmented Simulation (Entropic and Spatial Constraints)}

To examine how spatial and entropic factors influence viability, an augmented implementation introduces the extended penalty term $\Phi(t)$ and a resilience parameter $R_n$ that modulates the effect of structural redundancy.

The augmentation introduces an entropic-diffusive penalty $\Phi(t) = \alpha_S T\,\Delta S_{\text{local}}(t) + \beta_D \frac{R T}{D_0} D(x, t)$ and a resilience factor $R_n$. This preserves the baseline simulation structure while allowing persistence boundaries to shift with entropy gradients and diffusion rates.

\begin{algorithm}[H]
\caption{Augmented Persistence Simulation with Entropic and Diffusive Modifiers}
\begin{algorithmic}[1]

\State $S \gets 0$
\State $\mathcal{P} \gets \emptyset$

\For{each time-step $t$}

    \State $z \gets$ SampleEnvironmentalEnergy($t$)

    \For{each reaction chain $\mathcal{C}$}

        \State $E \gets z + S$
        \State $\mathcal{C}.\text{viable} \gets \text{True}$

        \For{each reaction $i$ in $\mathcal{C}$}

            \State $(x_i, r_i, s_i) \gets$ ReactionParameters($i$)

            \If{$E < x_i$}
                \State $\mathcal{C}.\text{viable} \gets \text{False}$
                \State \textbf{break}
            \EndIf

            \State $E \gets E - x_i + r_i$
            \State $S \gets S + s_i$

        \EndFor

        \If{$\mathcal{C}.\text{viable} = \text{False}$}
            \State $\mathcal{P} \gets \mathcal{P} \cup \{(t, \mathcal{C}, \text{False})\}$
            \State \textbf{continue}
        \EndIf

        \State $\Delta S_{\text{loc}} \gets$ ComputeLocalEntropyChange($\mathcal{C}, t$)
        \State $D \gets$ ComputeDiffusivity($\mathcal{C}, t$)
        \State $R \gets$ ComputeResilience($\mathcal{C}$)

        \State $\Phi \gets \alpha_S \cdot T \cdot \Delta S_{\text{loc}}
            \;+\; \beta_D \cdot (R_g T / D_0) \cdot D$

        \State $y_{\text{base}} \gets E$
        \State $y_{\text{aug}} \gets y_{\text{base}} + R - \Phi$

        \If{$y_{\text{aug}} \ge 0$}
            \State $\mathcal{C}.\text{viable} \gets \text{True}$
        \Else
            \State $\mathcal{C}.\text{viable} \gets \text{False}$
        \EndIf

        \State $\mathcal{P} \gets \mathcal{P} \cup \{(t, \mathcal{C}, \mathcal{C}.\text{viable})\}$

    \EndFor

    \State $S \gets S - \gamma S$

\EndFor

\State \Return $\mathcal{P}$

\end{algorithmic}
\end{algorithm}

\begin{figure}[h!]
\centering
\includegraphics[width=0.82\textwidth]{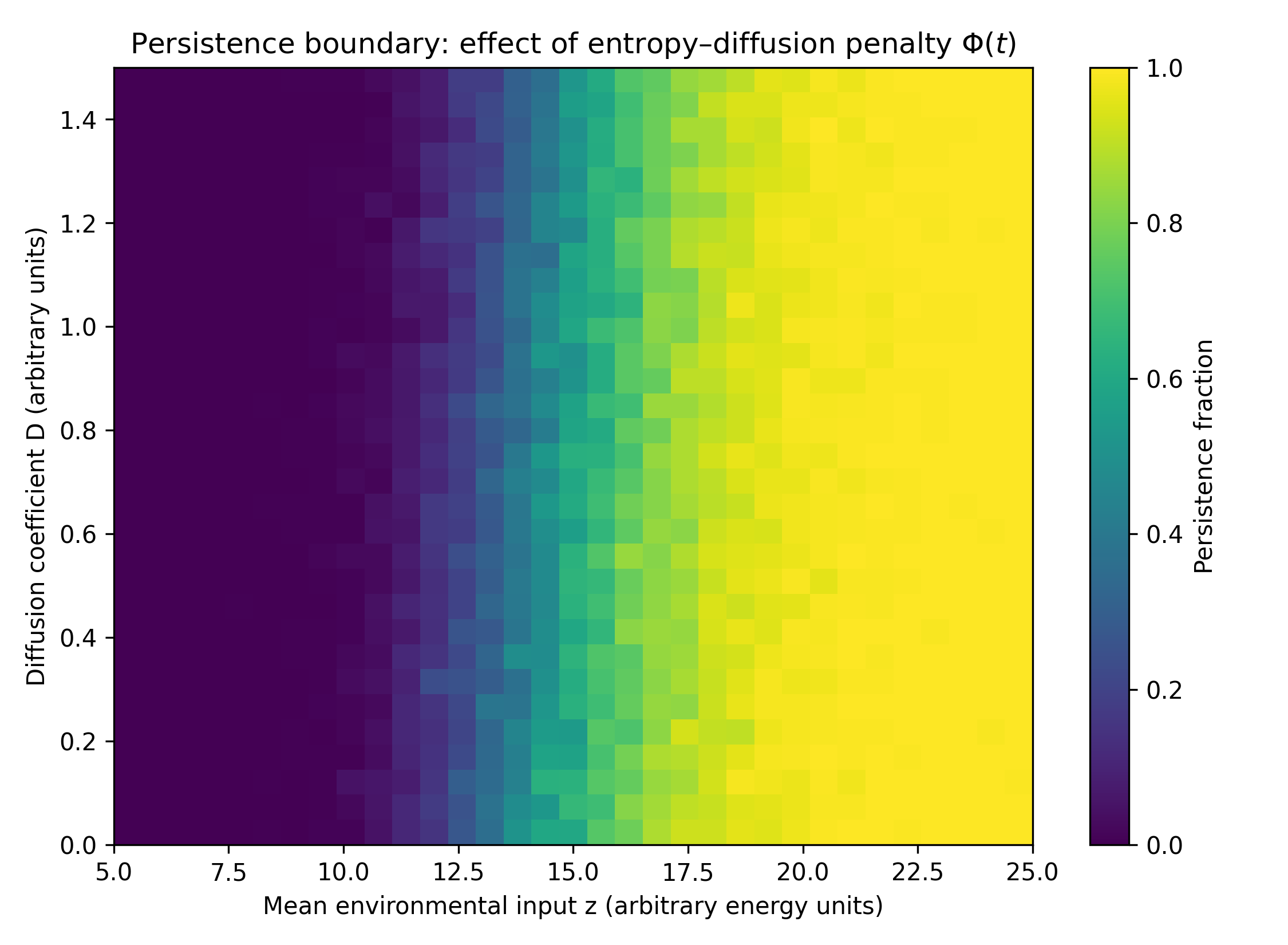}
\caption{Augmented persistence boundary including the entropic-diffusive penalty $\Phi(t) = \alpha / \Delta S_{\text{net}} + \beta D$. The heatmap shows the fraction of trials with $y'(t)\geq 0$ across intervals as a function of mean environmental input $\bar{z}$ and diffusion coefficient $D$ at fixed variability $\sigma_z$. Increasing $D$ raises the penalty and shifts the boundary to higher $\bar{z}$, illustrating how spatial dispersion can suppress persistence even under the same mean input.}
\label{fig:persistenceboundary_augmented}
\end{figure}
\newpage
\subsubsection{Persistence Landscape Construction}

To evaluate how variations in environmental energy availability influence the likelihood of reaction
persistence, the model computes a persistence landscape across a range of parameter combinations.
Each point in this landscape represents the proportion of time-steps for which the augmented
persistence condition $y'(t) \ge 0$ is satisfied under a particular environmental configuration.  
The procedure samples environmental parameters, runs multiple independent simulations using either
the baseline or augmented update rules, and records the frequency of viable intervals. By repeating
this process across a grid of parameter values, the resulting matrix forms a quantitative
representation of the persistence landscape, which serves as the basis for the heatmaps and phase
regions shown in the results. Algorithm~3 summarises the computation of this landscape.

\begin{algorithm}[H]
\caption{Persistence Landscape Generation}
\begin{algorithmic}[1]

\Require Set of environmental parameter combinations $\{\theta_k\}$ 
\Require Number of time-steps $N_{\text{steps}}$
\Require Number of independent runs per parameter set $N_{\text{runs}}$
\Require Simulation mode (\textsc{Baseline} or \textsc{Augmented})

\State Initialise landscape matrix $L \gets \emptyset$

\For{each parameter set $\theta_k$ in $\{\theta_k\}$}

    \State $\text{persistent\_intervals} \gets 0$
    \State $\text{total\_intervals} \gets 0$

    \For{$r \gets 1$ to $N_{\text{runs}}$}

        \State ConfigureEnvironment($\theta_k$)
        \State Initialise stored free energy $S \gets 0$

        \For{$t \gets 1$ to $N_{\text{steps}}$}

            \If{SimulationMode = \textsc{Baseline}}
                \State $\mathcal{P}_t \gets$ RunBaselineStep($t$, $S$)
            \Else
                \State $\mathcal{P}_t \gets$ RunAugmentedStep($t$, $S$)
            \EndIf

            \For{each record $(t, \mathcal{C}, \mathrm{viable})$ in $\mathcal{P}_t$}
                \State $\text{total\_intervals} \gets \text{total\_intervals} + 1$
                \If{$\mathrm{viable} = \text{True}$}
                    \State $\text{persistent\_intervals} \gets \text{persistent\_intervals} + 1$
                \EndIf
            \EndFor

        \EndFor

    \EndFor

    \State $f_{\text{persist}}(\theta_k) \gets 
        \text{persistent\_intervals} / \text{total\_intervals}$

    \State Store $f_{\text{persist}}(\theta_k)$ in $L$ at the coordinates corresponding to $\theta_k$

\EndFor

\State \Return landscape matrix $L$

\end{algorithmic}
\end{algorithm}

\noindent
\subsubsection*{Summary}

Taken together, the three simulation procedures provide a coherent computational counterpart to the
theoretical persistence framework. The baseline algorithm evaluates viability under fluctuating
environmental free-energy inputs, while the augmented algorithm incorporates the additional
entropic, diffusive, and structural contributions that shape the free-energy balance over time.
Algorithm~3 extends these procedures to construct persistence landscapes, enabling systematic
exploration of how different environmental and molecular parameters influence the frequency of
intervals in which the condition $y'(t) \ge 0$ is satisfied. These simulations therefore translate the
analytical model into quantitative predictions, allowing the behaviour of diverse hypothetical
reaction networks to be compared across parameter ranges and providing an empirical foundation for
the persistence maps presented in the results.  In subsequent sections, this computational foundation is generalized into the Thermodynamic Abiogenesis Likelihood Model (TALM), which extends these persistence criteria to broader chemical and planetary contexts.

\section{Thermodynamic Abiogenesis Likelihood Model (TALM)}
The Thermodynamic Abiogenesis Likelihood Model (TALM) applies the persistence inequality and its
modulators as an analytical framework for evaluating viability across chemical ensembles and planetary
environments. It generalizes the persistence inequality into a probabilistic framework that
incorporates entropic transformations, diffusion constraints, and spatial organization as modulators
of molecular persistence.

\subsection{Entropy-Driven Exploration of Chemical Space}

TALM adopts the view that in open, far-from-equilibrium systems, entropy production enables
exploration of configuration space: variable environmental energy fluxes increase global entropy
while permitting localized entropy reductions associated with transient structures and reaction
networks. Stochastic configurations generated in this way are then filtered by the reaction viability
inequality, linking entropy-driven exploration to thermodynamic selection.

\subsection{Reaction Viability in Structured Environments}

TALM integrates the augmented viability function:

\begin{equation}
y'(t) = y(t) + R(t) - \Phi(t)
\end{equation}

Here the components are interpreted as follows:

\begin{itemize}
  \item $y(t)$ is the underlying free-energy balance incorporating environmental input, stored free
  energy, and reaction-associated energetic terms
  \item $R(t)$ is the resilience contribution associated with structural or contextual stability
  \item $\Phi(t)$ is the entropic--diffusive penalty associated with maintaining local organization
\end{itemize}

The penalty term and resilience contribution are scaled using two constants:
\begin{itemize}
  \item $\alpha_S$ scales the entropy-related cost of maintaining local order
  \item $\beta_D$ scales the diffusive cost associated with spatial dispersion
\end{itemize}

The entropic--diffusive penalty is therefore written as
\[
\Phi(t) = \alpha_S\,T\,\Delta S_{\text{local}}(t)
        + \beta_D\,\frac{RT}{D_0}\,D(x,t),
\]
which captures the free-energy cost of counteracting thermal disorder and diffusion in
non-equilibrium environments. This formulation provides a dimensionally consistent representation aligned with the
augmented persistence model.

\subsection{Likelihood Formulation}

The likelihood that at least one interval supports persistence is

\begin{equation}
P_{\text{selective}} = 1 - \big(1 - P(y'(t) \ge 0)\big)^{N_{\text{eff}}},
\end{equation}

where $P(y'(t) \ge 0)$ is the probability that the free-energy balance is favourable during a sampled
interval, and $N_{\text{eff}}$ is the effective number of independent instances. In spatially or
temporally correlated settings, independence is reduced such that

\[
N_{\text{eff}} = \frac{N}{\,1 + 2 \sum_{\tau \ge 1} \rho(\tau)\,},
\]

with $\rho(\tau)$ representing the autocorrelation of viability across successive intervals. This
correction accounts for partial dependence between samples in confined or resource-limited
environments.

\subsection{Comparative Context: Earth and Icy Moons}

To illustrate the applied potential of the TALM framework, we consider its implications for three
planetary environments: early Earth, Europa, and Enceladus. These contexts exhibit the chemical
diversity, compartmentalisation opportunities, and variable energy inputs relevant to persistence
dynamics. These examples are illustrative rather than quantitative and demonstrate how the
persistence inequality may be used to compare environmental favourability rather than to provide
specific predictions.

\textbf{Early Earth} featured diurnal heating, geothermal activity, and intermittent hydration
intervals, producing confined microenvironments (e.g., clay layers, mineral pores) with reduced
diffusivity and variable free-energy supply, factors that increase the probability of $y'(t)\ge 0$
within TALM.

\textbf{Europa} may possess hydrothermal activity beneath an icy shell, with tidal flexing driving
sustained energy input. Although lacking surface cycling, the combination of geothermal gradients and
mineral interfaces could support moderate $P_{\text{selective}}$ values, particularly in porous or
compartmentalised regions.

\textbf{Enceladus} shows evidence of internal heating and organic-rich plumes, suggesting transient
microenvironments with spatial structure and chemical gradients. TALM predicts that even short-lived
favorable intervals could support molecular persistence where thermal or chemical pulses align with
free-energy requirements.

These examples illustrate how TALM provides a qualitative means of comparing abiogenesis potential
across planetary settings.

\subsection{Urability as a Prebiotic Filter}

Just as habitability describes environments where life can be sustained, \textit{urability} refers to
environments capable of supporting the origin of life. It characterizes the prebiotic window in which
physical and chemical conditions---such as compartmentalisation, accessible substrates, and
non-equilibrium free-energy flux---enable the emergence of molecular complexity prior to heredity or
replication \cite{deamer2021urability}.

Within the TALM framework, urability functions as a pre-filter for identifying planetary settings
where preconditions for abiogenesis are plausible. TALM then evaluates whether reaction networks in
those settings satisfy the persistence criterion with sufficient frequency. Together, urability and
TALM provide a complementary two-tier approach to assessing the likelihood of abiogenesis.

\section{Illustration of the Model}

\subsection{Reaction chain model}

\subsection{Energy dynamics over time}

\begin{figure}[htbp]
\centering
  \includegraphics[width=1\textwidth]{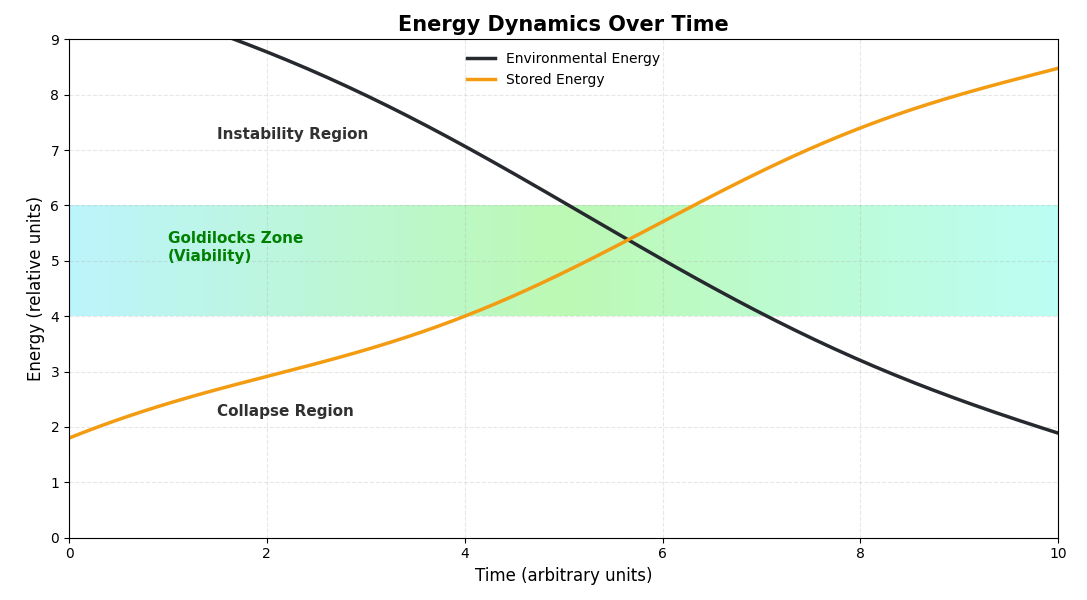}
  \caption{The energy–time plot illustrates how environmental energy input and internal energy storage jointly determine the viability of a reaction sequence across fluctuating conditions.}
  \label{fig:energy_dynamics}
\end{figure}

Figure~\ref{fig:energy_dynamics} illustrates energy dynamics over time, showing the interaction between stored energy and energy used in reactions.

\newpage
\subsection{Numerical Example: Reaction Chain Viability}
To illustrate the viability inequality in practice, we consider a hypothetical reaction chain consisting of three sequential reactions, each with defined energy requirements ($x_i$), energy releases ($r_i$), and stored energy ($s_i$). We evaluate the chain's persistence under two different environmental energy conditions, represented by $z(t)$.

\subsubsection{Reaction Chain Parameters}
We define the following reactions:

\begin{center}
\begin{tabular}{|c|c|c|c|}
\hline
\textbf{Reaction} & $x_i$ (Required) & $r_i$ (Released) & $s_i$ (Stored) \\
\hline
$R_1$ & 8 & 2 & 1 \\
$R_2$ & 5 & 3 & 2 \\
$R_3$ & 7 & 1 & 1 \\
\hline
\end{tabular}
\end{center}

The system attempts to perform all three reactions sequentially, starting with energy $z(t)$ from the environment.

\subsubsection{Case 1: Sufficient Environmental Energy (\texorpdfstring{$z(t) = 10$})}
\begin{itemize}
  \item $R_1$: $z = 10 \rightarrow z - x_1 + r_1 = 10 - 8 + 2 = 4 \rightarrow$ Store $s_1 = 1$
  \item $R_2$: $z = 4 \rightarrow z - x_2 + r_2 = 4 - 5 + 3 = 2 \rightarrow$ Store $s_2 = 2$
  \item $R_3$: $z = 2 \rightarrow z - x_3 + r_3 = 2 - 7 + 1 = -4 \rightarrow$ Insufficient energy
\end{itemize}
\textbf{Result:} System collapses at step 3.

\subsubsection{Case 2: Higher Energy Input (\texorpdfstring{$z(t) = 15$}{z(t) = 15})}
\begin{itemize}
  \item $R_1$: $z = 15 \rightarrow 15 - 8 + 2 = 9 \rightarrow$ Store $s_1 = 1$
  \item $R_2$: $z = 9 \rightarrow 9 - 5 + 3 = 7 \rightarrow$ Store $s_2 = 2$
  \item $R_3$: $z = 7 \rightarrow 7 - 7 + 1 = 1 \rightarrow$ Store $s_3 = 1$
\end{itemize}
\textbf{Result:} Chain completes successfully. System persists and stores cumulative energy for future use.

This simplified example demonstrates the core logic of the model: system viability is governed not just by initial energy input but also by how reactions release or store energy internally. Even efficient systems may collapse under insufficient environmental support. Conversely, well-aligned chains survive and accumulate structural energy, potentially leading to selection pressure in energy-variable environments.

\subsection{Illustrative Example: Thermal Cycling on Early Earth}

To illustrate the applicability of the persistence framework under real environmental conditions, consider empirical measurements by Deamer and Damer (2020), who demonstrated that dehydration-rehydration intervals in hydrothermal fields drive non-enzymatic polymerization of nucleotide-like molecules under prebiotic conditions by recorded daily temperature fluctuations ranging from 30-90\textdegree C in surface hydrothermal pools at Kamchatka\cite{deguzman2014oligonucleotides} \cite{ross2019drywet}. These fluctuations correspond to molecular-scale energy variations on the order of \(10^{-20}\) to \(10^{-18}\) joules per molecule making ideal testbeds for modeling \(z(t)\) the time-dependent environmental energy input.

We define \(z(t)\) as a sinusoidal function representing periodic thermal cycling:

\[
z(t) = z_{\max} \cdot \sin(\omega t) + z_{\text{offset}}
\]

Where:
- \(z_{\max}\) is the peak energy input during daytime,
- \(\omega\) is the angular frequency corresponding to a full day-night cycle,
- \(z_{\text{offset}}\) accounts for geothermal heat or atmospheric energy retention during cooler periods.

During daytime, energy peaks may support energetically demanding reactions (\(\sum x_i\) is high), while nighttime introduces an energy deficit. Only systems with sufficient internal storage \(S(t)\), or inherently low energy requirements \(x_i\), will remain viable during low-input phases.

For example, let a reaction system require \(x_i = 5\) energy units to maintain structural viability and release \(r_i = 1\). If \(z(t)\) peaks at 6 and falls to 1 at night, then persistence through a full cycle is only possible for systems that can retain at least \(S(t) \geq 3\) units of stored energy from the high-input phase.

\begin{figure}[htbp]
    \centering
    \includegraphics[width=0.9\textwidth]{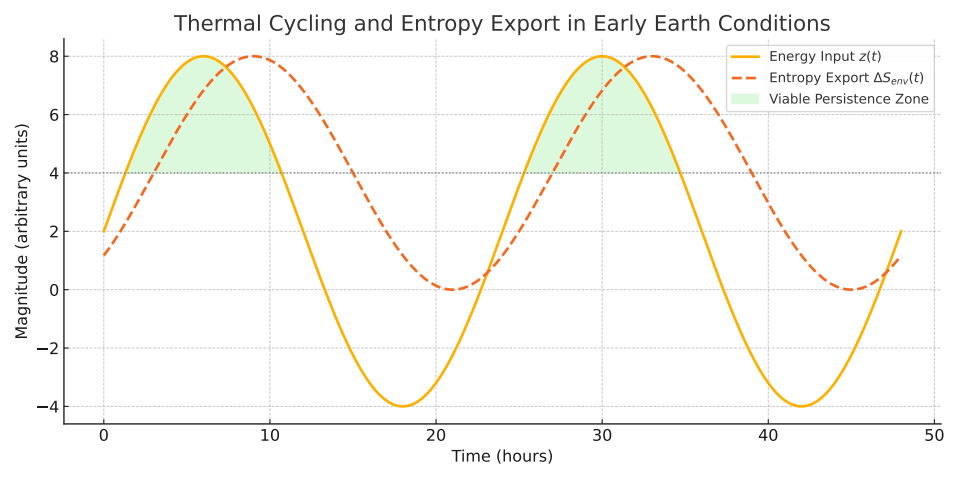}
    \caption{
        Simulated thermal cycling and entropy export in early Earth conditions. 
        The solid curve represents energy input $z(t)$ over a 48-hour period, modeled as a sinusoidal day-night cycle with baseline energy. 
        The dashed curve shows entropy exported to the environment, $\Delta S_{env}(t)$, phase-shifted to reflect thermal dissipation lag. 
        The shaded region indicates the viable persistence zone, where energy input exceeds the threshold required for reaction continuity. 
        This illustrates how persistent systems must not only align with fluctuating $z(t)$, but also export sufficient entropy to remain thermodynamically viable.
    }
    \label{fig:early_earth_entropy}
\end{figure}
This scenario demonstrates the central principle behind the persistence model: differential energy storage enables selective retention. Systems capable of buffering or storing energy during abundance become preferentially stable across environmental fluctuations. Thermal cycling, when combined with compartmentalization and diffusion-limiting substrates, thus constitutes a plausible natural engine for prebiotic selection.

Furthermore, such persistence would only be thermodynamically viable if local decreases in entropy,  associated with maintaining molecular structure across intervals,  are offset by corresponding entropy export to the environment. In this way, day-night thermal cycling could support transient self-organization that satisfies the second law, especially in compartmentalized or surface-adhered systems where diffusion is constrained.  This type of modeling could be extended to other environments with time-variable or localized energy gradients, such as hydrothermal vents.

\subsection{Illustrative Application to Planetary Environments}

Although this work focuses primarily on early Earth and hydrothermal field settings, the framework is not restricted to terrestrial conditions. In principle, any planetary or lunar environment that can be characterized in terms of (i) energy input statistics $z(t)$, (ii) chemical diversity and availability of reactive substrates, and (iii) the presence or absence of diffusion-limiting structures can be mapped onto the persistence inequality and its augmented form.

A practical application would proceed in two stages. First, an environment would be assessed for \textit{urability} in the sense of Deamer and colleagues \cite{deamer2021urability}: whether it plausibly offers fluctuating energy sources, suitable solvents, and some means of spatial confinement. Second, for environments that meet these baseline criteria, the persistence model could be utilized to sample ensembles of reaction networks under estimated $z(t)$ profiles and diffusion constraints, yielding comparative values of $P_{\text{selective}}$ across candidate worlds. This would not provide a literal probability that life exists, but rather a physically grounded ranking of environments by their propensity to support persistent, energy-compatible reaction networks.

A full implementation of such a planetary survey lies beyond the scope of the present paper, but the framework outlined here is expressly designed to permit this type of extension once appropriate environmental and geochemical constraints are available.

\section{Theoretical Context and Limitations}
The framework presented in this paper aligns with and extends a body of interdisciplinary work exploring the thermodynamic and information-theoretic basis for the origin of life. While traditional models of evolution emphasize replication and heredity, our approach emphasizes the prior condition necessary for such processes: the persistence of reaction systems under energetic constraints.

\subsection{Limitations of the Model}

\subsubsection{Simplified Chemical Representation}

The model treats reactions as discrete energetic steps parameterized by $x_i$, $r_i$, and $s_i$. This abstraction omits mechanistic detail—such as intermediates, catalysis, and solvent-structure couplings—in favor of generality. Although this enables broad applicability, its mapping to specific prebiotic chemistries must be established empirically. Experimental and computational studies will be required to determine realistic parameter ranges for early Earth or analogous environments.

\subsubsection{Environmental Energy Abstraction}

The environmental input term $z(t)$ is modeled generically to encompass diverse prebiotic energy sources (thermal cycling, photonic flux, geothermal gradients). While appropriate for a general framework, it does not encode any specific planetary setting. Future work could incorporate empirical energy profiles—diurnal intervals, geothermal temperature curves, or seasonal variations—either analytically or through simulation. Extending $z(t)$, $x_i$, and $r_i$ to depend explicitly on temperature or other environmental variables (e.g., Arrhenius-like behavior) would allow more detailed comparisons without altering the framework’s core structure.

\subsubsection{Stored Energy and Buffering}

The model treats stored energy as a structural or chemical reserve capable of buffering fluctuations in $z(t)$. This captures the intuitive idea that systems with internal energy stores can survive transient loss events. However, the biophysical mechanisms that might instantiate $S(t)$—such as polymerized energy carriers or metastable structures—remain to be explored in real chemical systems. Quantifying the magnitude and decay of such stores is an important next step for connecting the inequality to experiment.

\subsubsection{Entropy and Spatial Extensions}

The augmented inequality incorporates entropic and spatial terms ($\Delta S_{local}$, $\Delta S_{env}$, $D(x,t)$), enabling treatment of self-assembly, diffusion-limited organization, and compartmentalization. These extensions broaden the model’s physical realism but also introduce additional parameters that must be experimentally constrained. The current formulation treats these contributions phenomenologically rather than mechanistically, and follow-up work will be required to map these abstract terms onto measurable quantities in specific chemical systems.

\subsubsection{Reaction Networks and Topology}

The present implementation considers linear reaction chains, whereas real prebiotic systems likely involved branching pathways, redundancy, and feedback. Extending the framework to network topology would clarify how resilience emerges from alternative pathways and whether autocatalytic or DKS-like organizations appear naturally within the persistence criterion. This represents a natural avenue for future generalization.

\subsubsection{Thermodynamic Origins of Selection}

The model posits that selection precedes heredity: structures that maintain viability under fluctuating energy inputs persist, while others collapse. This thermodynamic filtering aligns with known features of dissipative systems, autocatalytic organization, and energy-entropy-compatible self-assembly. The resulting picture is probabilistic rather than discrete: abiogenesis emerges as a gradual transition in which open systems explore chemical space while energy flow retains those configurations most compatible with the local environment.

These persistence inequalities therefore offer a pre-replicative substrate on which later Darwinian constraints—such as those formalized by Kolchinsky—might act once replication appears. Verifying the predictions made here would establish whether energy-structure coherence provides the physical preconditions for natural selection, bridging the gap between purely thermodynamic organization and the efficiency-driven competition characteristic of life.

\subsection{Related Works and Theories}

\subsubsection{Autocatalytic Sets and Organizational Closure}
Stuart Kauffman's theory of autocatalytic sets \cite{kauffman1993origins} laid the foundation for viewing life as a network of reactions that sustain and propagate themselves. His work framed emergence not as a consequence of genetic control but as a product of combinatorial chemistry and mutual reinforcement. Our model complements this by suggesting that not all autocatalytic chains persist equally,  only those aligned with dynamic energy availability survive.

\subsubsection{Dissipative Structures and Far-From-Equilibrium Systems}
Ilya Prigogine's concept of dissipative structures \cite{prigogine1977self} posits that ordered systems can arise spontaneously in non-equilibrium environments through energy dissipation. Our framework draws from this principle by proposing that prebiotic systems capable of balancing energetic inflow and reaction requirements naturally self-stabilize, forming precursors to selection mechanisms.

\subsubsection{Wet–dry cycling and non-enzymatic polymerization}

Song et al.'s work on nucleotide polymerization \cite{song2012wetdry} investigates how nucleic acid monomers, when subjected to repeated wet-dry cycles under conditions simulating volcanic hot springs, can spontaneously polymerize into long chains. This study demonstrates that RNA and DNA-like oligomers can form in the absence of enzymes, highlighting a plausible non-enzymatic pathway for early genetic polymers. Their findings support the idea that environmental cycling, particularly dehydration-rehydration dynamics, can drive condensation reactions that would otherwise be thermodynamically unfavorable in continuously aqueous conditions.

\subsubsection{Generation of oligonucleotides under hydrothermal conditions by non-enzymatic polymerization}
Deguzman et al's work on hydrothermal conditions
\cite{deguzman2014oligonucleotides} studies prebiotic synthesis under fluctuating hydration that support the idea of nonequilibrium thermodynamic cycling can create and stabilize chemical complexity, without the need for enzymes or genetic templates.

\subsubsection{Dissipative Adaptation}
Jeremy England's work on dissipative adaptation \cite{england2013statistical} formalized how driven systems statistically evolve to dissipate energy more efficiently. We align with this view, proposing that selection arises not from genetic memory, but from energy-aligned structural persistence, chains that survive fluctuations in $z(t)$ naturally dominate the prebiotic landscape.

\subsubsection{Dynamic Kinetic Stability}
Addy Pross's theory of dynamic kinetic stability \cite{pross2012what} reframes life's origin as a transition from thermodynamic to kinetic dominance,  that is, from stable molecules to stable replicative processes. Our model precedes this transition, exploring the energetic logic by which certain reaction chains become stable enough to eventually support replication.

\subsubsection{The Hydrophobic Effect and Amphiphilic Self-Assembly}

The hydrophobic effect explains the spontaneous organization of amphiphilic molecules in aqueous environments. First formalized in detail by Charles Tanford \cite{tanford1980hydrophobic}, the effect arises from water's tendency to exclude non-polar groups in order to preserve its hydrogen-bonding network. When hydrophobic molecules or molecular segments are introduced into water, they disrupt the local structure of the solvent, resulting in an entropic penalty. To minimize this disruption, water reorganizes itself into a more ordered shell around the solute. However, when multiple hydrophobic groups cluster together, the overall surface area exposed to water is reduced, thereby releasing structured water molecules back into the bulk and increasing the total entropy of the system.

This process underlies the self-assembly of amphiphilic molecules-those with both hydrophilic and hydrophobic domains, into micelles, bilayers, and vesicles. The hydrophobic tails aggregate to minimize exposure to water, while the hydrophilic heads remain solvent-accessible. As a result, structurally coherent assemblies emerge without any external guiding information.

\paragraph*{Hydrophobic Theories of Self-Organization}

A number of theoretical models have been developed to explain the behavior of amphiphilic molecules and the emergence of biologically significant structures such as micelles, bilayers, and folded proteins, based on hydrophobic interactions.

\subparagraph*{Micelle formation}
Micelle formation represents a foundational example of hydrophobic self-assembly. Amphiphilic molecules such as fatty acids or detergents spontaneously aggregate in aqueous environments to form micelles-spherical structures in which the hydrophobic tails are shielded from water by an outer layer of hydrophilic headgroups. This process minimizes the entropic cost of disrupting water's hydrogen-bonding network. 

Thermodynamic models, such as those developed by Maibaum et al (2004), have demonstrated how micelle formation can be simulated using coarse-grained particles interacting via effective hydrophobic potentials \cite{maibaum2004micelles}. Their work provides a statistical mechanical basis for the emergence and reorganization of micelles under different thermodynamic constraints. However, these models generally assume that the amphiphilic molecules are already present in optimal concentrations and that micelle formation is driven purely by equilibrium considerations.

\subparagraph*{Lipid Bilayers and Bilayer Formation}
  Brannigan et al (2005). introduced an implicit-solvent model in which coarse-grained lipid-like amphiphiles self-assemble into bilayer structures \cite{brannigan2005bilayers}. Their work captures membrane fluidity and mechanical flexibility through effective hydrophobic interactions, providing a simplified physical basis for bilayer formation. However, their model begins with designed molecules and assumes favorable interactions, rather than offering an explanation for why such structures would emerge prebiotically. Hence the thermodynamic persistence model can be viewed as a complementary framework, offering criterion by which certain amphiphilic molecules, among many possible prebiotic variants, would be preferentially retained under fluctuating energetic conditions.

\subparagraph*{Interface-Sensitive Assembly Behavior}
Patel et al (2011). extended the Lum-Chandler-Weeks theory to show how hydrophobic effects at extended surfaces can catalyze assembly and influence water structure \cite{patel2011hydrophobic}. This framework contributes a deeper understanding of how surface geometry and water fluctuations modulate hydrophobic assembly. These environmental variables could be integrated into the viability model to evaluate how structured surfaces affect reaction persistence, pointing to a possible bridge with interface-sensitive assembly behavior.

\subparagraph*{The Energy Funnel Model of Protein Folding}
Hummer et al (1998) developed a statistical mechanical theory for hydrophobic solvation at the molecular scale \cite{hummer1998hydrophobic}, revealing how entropy and solvent structuring contribute to protein folding. This was built on to propose the energy funnel model of protein folding \cite{dill1997levinthal}, resolving the Levinthal paradox by showing how hydrophobic collapse drives efficient exploration of conformational space. The persistence inequality model does not directly reference model folding, but it may explain why certain polymers with folding-prone sequences might persist longer than others, thus serving as a precondition for the emergence of foldable biomolecules.

\subsection{Novelty of the Present Work}

While these foundational theories address persistence, emergence, and order, they rarely formalize persistence in terms of a reaction viability inequality under dynamically shifting energy constraints.  Existing frameworks – from dissipative structure theory \cite{Prigogine1972} to maximum entropy production principles \cite{Swenson1989} and dynamic kinetic stability \cite{pross2012what} - successfully describe how organized states can arise and be maintained in driven environments. However, these approaches typically operate at macroscopic or statistical levels and do not provide a local, quantitative criterion for determining whether a specific prebiotic reaction system will continue, collapse, or transition under fluctuating external conditions. The present work introduces such a criterion in the form of a reaction viability inequality that links environmental energy input, internal energetic costs, and entropy-related penalties to the time-dependent persistence of a reaction network.

This inequality-based formulation is distinct in three respects. First, it identifies selection-like behaviour \emph{prior to} heredity or replication by showing how differential persistence emerges purely from thermodynamic compatibility with the driving environment. This contrasts with frameworks such as England’s dissipative adaptation, which presuppose replication-like transitions \cite{england2013statistical}. Second, it provides a minimal but generalizable filter that applies at the level of individual reactions, small networks, or mesoscale assemblies, complementing global dissipation-optimization views that operate at much larger scales, such as biospheric entropy production models \cite{Ulanowicz1987}. Third, it translates the thermodynamic constraints on persistence into experimentally testable conditions, enabling direct evaluation in systems such as wet–dry polymerization cycles and lipid-based protocells where environmental driving determines structural longevity \cite{Deamer2017}.

In this way, the framework does not attempt to replace existing non-equilibrium theories, but instead provides a local, mechanistic bridge between environmental driving and the differential survival of reaction networks – a domain where current formalisms offer descriptive insight but no operational criterion for viability.

\section{Implications, Predictions, and Falsifiability}
\subsection{Theoretical Outlook}

The integration of entropy production, diffusion constraints, and spatial structure into the persistence model has clarified how thermodynamic forces can shape the emergence of chemical order in prebiotic contexts. By reframing entropy not as an antagonist to organization, but as a driver of exploration within chemical state space, we offer a new lens through which the formation of persistent molecular configurations can be understood.

The framework presented, provides a means of translating this conceptual model into a probabilistic structure applicable across diverse environmental conditions. The reaction viability inequality, extended via $y'(t)$, allows for the estimation of persistence under spatial and energetic modulation. This yields not only a theoretical basis for selection-like behavior in prebiotic chemistry but also a potential comparative lens across planetary systems.

Continued refinement of this approach, through simulation of reaction networks, laboratory analogs of compartmentalized cycling systems, and the thermodynamic profiling of planetary bodies, may illuminate the constraints and pathways most conducive to the emergence of persistent reaction architectures. In this light, the framework serves not only as a theoretical advance but as a scaffold for the systematic comparison of environments where the thermodynamic roots of natural selection may have taken hold.

Moreover, pairing the model with Urability, the environmental pre-screening framework \cite{deamer2021urability}, which assesses whether a planetary setting possesses the thermodynamic and chemical conditions necessary for prebiotic complexity, provides a two-tiered approach by first evaluating urability, then applying the Thermodynamic Abiogenesis Likelihood Model. This provides a robust, scalable method for identifying environments most conducive to abiogenesis.

The expanded test case scenarios now incorporated into this work reinforce the models generality. They demonstrate that the persistence inequality can be applied not only to abstract reaction networks, but to real physical systems, such as folding oligomers, membrane dynamics, and compartmental reactions. These testable extensions highlight the Thermodynamic Abiogenesis Likelihood Model predictive versatility and provide a concrete roadmap for future validation. Crucially, they show that persistence inequality is not a metaphorical selection mechanism, but a quantitative framework grounded in measurable energy flows.

Ultimately, this model invites us to reconsider life's origin not as a singular event but as a recurring possibility, a statistical consequence of energy flow, structure, and entropy at work within the laws of physics.

\subsection{Predictions and Falsifiability}

The experimental and simulation scenarios described in Sections 8-10 collectively serve a single purpose: to determine whether the reaction viability inequality and its augmented persistence function accurately describe how chemical systems maintain structure under fluctuating energy input. Each test therefore represents an opportunity for the model to be proven wrong.

\subsection{Purpose of the suite}
Together, the tests probe three linked claims of the thermodynamic persistence inequality model:

\begin{enumerate}
    \item Persistence is governed by energetic alignment, not replication;
    \item Entropy export and spatial confinement are the decisive modulators of structural stability; and
    \item These same principles scale from molecular assemblies to planetary environments.
\end{enumerate}

\subsection{Test Scenarios}
\subsubsection{Entropy-Driven Self-Assembly in Amphiphilic Systems}

A tractable and highly illustrative system for testing the persistence inequality model involves the self-assembly behavior of amphiphilic molecules in water, where structure formation is driven primarily by the hydrophobic effect. This effect arises because water molecules form hydrogen-bonded networks. When non-polar (hydrophobic) molecules are introduced, they disrupt these networks, lowering the system's entropy. Water responds by ordering tightly around the intruding molecules. However, when many hydrophobic molecules are present, they tend to cluster together, freeing the surrounding water to return to a disordered state, thus increasing entropy.

To explore this within the persistence model framework, consider a series of aqueous systems each containing 18 mL of water (1 mole) and 0.01 mole of one of three solutes: sodium acetate, sodium octanoate, and sodium dodecanoate. These molecules differ only in the length of their hydrophobic tails. When air is gently introduced, observable differences arise. Sodium acetate, which lacks a hydrophobic tail, dissolves completely with no visible change. Sodium octanoate, with a moderate tail length, forms a transient bubble. Sodium dodecanoate, with a longer hydrophobic chain, forms a stable bubble and, under microscopic inspection, many vesicles.

These behaviors are predicted as a function of chain length and the system's ability to manage entropy and energy flow. The longer tails drive stronger self-association, lowering diffusion ($D(x,t)$) and increasing structural resilience ($R_n$), while also allowing water to release more entropy as structured cages collapse. Thus, the entropy cost of forming vesicles is more than offset by entropy export to the environment:
\[
\Delta S_{net} = \Delta S_{env} - \Delta S_{local} > 0
\]
leading to a reduced $\Phi(t)$ and positive $y'(t)$.

We predict that structural persistence-bubbles, micelles, and vesicles emerge at a critical chain length $L_c$. Based on this system:
\begin{itemize}
  \item For $L < C8$, self-assembly is minimal; no structures persist.
  \item At $L \approx C10$, vesicle formation becomes likely.
  \item Between $C10$ and $C14$, vesicles are increasingly stable.
  \item Beyond $C14$, solubility may decrease, limiting structure formation.
\end{itemize}

\paragraph*{Tests}
\begin{itemize}
  \item \textbf{Context/Gap:} Experimental studies show that fatty acid vesicles depend strongly on chain length for their stability, but no general thermodynamic criterion has been formulated for identifying the thresholds of persistence under fluctuating conditions.
  \item \textbf{Model Action:} The thermodynamic persistence model translates amphiphile self-assembly into a balance between energy thresholds for aggregation and entropy penalties from solubility and dispersion.
  \item \textbf{Prediction (Expectation):} Vesicles shorter than C8 will fail to persist, vesicles near C10 will form but remain fragile, vesicles between C10 and C14 will persist more reliably, and chains longer than C14 will become limited by poor solubility.
  \item \textbf{Falsification:} If experiments show stable vesicles persisting well below C8 or well above C14 without confinement or catalysis, or if vesicle stability is largely independent of chain length, the framework would be falsified in this domain.
\end{itemize}

\subsubsection{Surface-Catalyzed Reaction Viability}

Mineral surfaces can influence the persistence of prebiotic reaction networks by modulating local energy dissipation, adsorption, and spatial retention. Within the persistence framework, the viability of a reaction sequence is evaluated using the condition
\[
y^{\prime}(t) = y(t) + R_n - \Phi(t) > 0 .
\]
Surface properties influence the loss term $\Phi(t)$ by altering water activity, reactant residence time, and the extent of molecular dispersal. Hydrophobic or structurally repetitive surfaces may restrict diffusion and lower environmental loss, whereas highly hydrated or reactive surfaces may increase $\Phi(t)$ through accelerated hydrolysis or rapid desorption.

To test this, we propose placing identical chemical mixtures on two surfaces: one smooth and water-repelling (hydrophobic), the other patterned and water-attracting (hydrophilic). Both are exposed to repeated wet-dry and heat intervals. If the prediction holds, the surface that supports a better energy balance, where \( y'(t) > 0 \), will see the reactions last longer.

This experiment would aim to demonstrate that even before enzymes or catalysts existed, the shapes and chemistry of early surfaces may have influenced chemical persistence, selecting not by function, but by physics.

The persistence inequality model predicts that structured surfaces will differentially support reaction persistence based on how they modulate energy dissipation:

\begin{itemize}
  \item On hydrophilic patterned surfaces: high water structuring increases \( \Phi(t) \), reducing persistence for some reactions.
  \item On hydrophobic flat surfaces: clustering may lower \( \Phi(t) \), favoring reactions with minimal net energy waste.
  \item Reaction networks aligned with local water organization will recur more reliably across energy intervals.
  \item Surfaces that buffer thermal or hydration fluctuations will increase \( y(t) \) enhancing persistence.
\end{itemize}

\paragraph*{Tests}
\begin{itemize}
\item\textbf{Context/Gap:} Mineral and rock surfaces are widely invoked in prebiotic chemistry, yet existing models often treat them as passive scaffolds rather than active thermodynamic constraints on persistence.

\item\textbf{Model Action:} In the thermodynamic persistence model, surface effects are captured by the entropic-diffusive penalty term $\Phi(t)$, which reflects water structuring, clustering, and buffering of fluctuations.

\item\textbf{Prediction (Expectation):} Hydrophilic patterned surfaces should increase $\Phi(t)$ and reduce persistence, while hydrophobic surfaces should reduce $\Phi(t)$ and favor persistence. Surfaces that buffer hydration or thermal intervals should increase $y(t)$ and enhance persistence.

\item\textbf{Falsification:} If persistence is observed to be unaffected by surface hydrophobicity, structuring, or buffering capacity, with systems persisting equally across all surface types, then the framework would be falsified in this context.
\end{itemize}
\subsubsection{Folding Bias in Random Oligomer Pools}

Non-templated peptides and RNA oligomers produced in prebiotic settings would have exhibited wide variability in structural stability. Many sequences would remain extended and solvent-exposed, rendering them susceptible to hydrolysis, thermal degradation, or dispersal. Others would spontaneously adopt compact folds that reduce the effective surface area exposed to the environment, particularly by burying hydrophobic residues or bases within an interior core.

Within the thermodynamic persistence framework, structural compaction contributes directly to viability through its effect on the persistence inequality:
\[
y'(t) = y(t) + R_n - \Phi(t) > 0
\]
Here, \( y(t) \) represents energy stored or buffered, folding adds to this; \( R_n \) is the net energy gained from any reaction (possibly zero here); and \( \Phi(t) \) is energy lost to the environment through heat or degradation.

A folded molecule might survive better simply because it reduces \( \Phi(t) \) and maintains \( y(t) \), keeping the total \( y'(t) \) positive.

To test this idea, we can create a large batch of random RNA or peptide sequences in water. Then we expose them to simulated early-Earth conditions: heating and cooling, drying and re-wetting, maybe even light exposure. After several intervals, we check which sequences are still around.

The model predicts that many of the survivors will be sequences that naturally fold, those that form compact, protected shapes in water. Folding, in this case, becomes a kind of passive survival mechanism. This would suggest that even before biology had evolved to \textit{select} useful sequences, nature was already filtering out unstable ones, simply based on their energy behavior in a rough environment.

Hence, foldable sequences will be statistically enriched among survivors in fluctuating environments:

\begin{itemize}
  \item Sequences that form hydrophobically collapsed folds exhibit higher  \( y(t) \) (energy retention).
  \item Folding reduces exposure to destructive fluctuations, lowering \( \Phi(t) \).
  \item Non-folding sequences degrade or hydrolyze more quickly during thermal or hydration cycling.
  \item Over time, the population shifts toward structurally compact, energy-buffering polymers.
\end{itemize}

\paragraph*{Tests}
\begin{itemize}
\item\textbf{Context/Gap:} Prebiotic polymerization would have produced vast numbers of sequences, most of which are non-folding. Existing theories have not explained how foldable sequences could be selectively retained prior to heredity.
\item\textbf{Model Action:} Folding alters thermodynamic viability by raising $y(t)$ (through energy retention in compact states) and lowering $\Phi(t)$ (by reducing exposure to destructive fluctuations).

\item\textbf{Prediction (Expectation):} Foldable polymers should persist more reliably across intervals of hydration and heating, while non-foldable polymers should degrade or hydrolyze more rapidly. Over time, foldable sequences should be statistically enriched.

\item\textbf{Falsification:} If experiments on polymer libraries reveal no persistence differences between foldable and non-foldable sequences under repeated cycling, the framework would be invalidated in this domain.
\end{itemize}
\subsubsection{Micelle-Embedded Reaction Selection}

Amphiphilic molecules in aqueous solution can self associate into micelles from what is referred to as the hydrophobic effect, where water avoids oily parts of molecules.  This process decreases the solvent exposed surface area and produces a protected hydrophobic core. Such structures can buffer environmental fluctuations by reducing the entropic and diffusive loss term $\Phi(t)$ and by stabilizing stored energy contributions to $y(t)$.

According to the persistence model, a micelle whose internal chemistry balances energy well, absorbing, storing, and releasing it in sync with the environment, should persist longer than one filled with wasteful or unstable reactions. The principle is summarized in the persistence inequality equation:
\[
y'(t) = y(t) + R_n - \Phi(t)
\]
Where \( y(t) \) is the energy held inside the micelle (e.g., as chemical potential), \( R_n \) is the net energy from internal reactions, and \( \Phi(t) \) is energy dissipated (through heat or structural strain).

Micelles that keep \( y'(t) > 0 \), that is, they maintain more usable energy than they lose, should survive longer through intervals of drying, heating, or stress.

To test this, scientists could create two groups of micelles with the same membranes but different contents: one group with gentle, energy-buffered reactions inside, and one with inefficient, chaotic reactions. After exposing both to intervals of dehydration and rehydration, the persistence inequality model predicts that the first group will re-form more often and persist longer.

We predicts that micelles with internal reactions that support energy alignment will dominate over time:

\begin{itemize}
  \item Encapsulated reactions with steady, buffered energy flow increase  \( y(t) \) and reduce internal \( \Phi(t) \).
  \item Micelles containing inefficient or highly dissipative reactions rupture or fail to reform.
  \item Micelle populations with aligned internal networks grow more numerous over intervals.
  \item Selection acts on the reaction-micelle composite, not just the structure.
\end{itemize}

\paragraph*{Tests}
\begin{itemize}
\item\textbf{Context/Gap:} Micelles are known to form spontaneously, but the impact of internal chemistry on their persistence under fluctuating conditions remains unclear.
\item\textbf{Model Action:} The thermodynamic persistence model incorporates encapsulated reactions into the viability inequality by coupling their energy dynamics to the host structure, modifying both $y(t)$ and $\Phi(t)$.

\item\textbf{Prediction (Expectation):} Micelles containing buffered internal reactions should maintain higher $y(t)$ and lower $\Phi(t)$, while micelles hosting dissipative reactions should rupture or fail to reform. Over time, selection will act on the reaction-micelle composite.

\item\textbf{Falsification:} If micelles persist equally regardless of whether they contain energy-aligned or dissipative reactions, the framework’s predictions would be contradicted.
\end{itemize}
\subsubsection{Bilayer Reconfiguration Under Cycling Energy Input}

Lipid bilayers can be modeled and simulated to understand how their structure depends on molecular properties like chain length and flexibility. The persistence model adds a new layer to this understanding: it suggests that bilayers which better regulate energy flow, absorbing and releasing it without collapsing, will persist longer under environmental cycles.

We express this idea by applying the simple inequality equation:
\[
y'(t) = y(t) + R_n - \Phi(t) > 0
\]
Here, \( y(t) \) is energy retained in the bilayer system (e.g., mechanical stability, membrane potential); \( R_n \) is energy added by any internal reactions or interactions; and \( \Phi(t) \) is energy lost through heat, rupture, or stress.

A bilayer that bends without breaking, or slows the flow of ions without fully blocking them, may minimize \( \Phi(t) \) and help keep \( y'(t) \) positive, even in a volatile environment.

To test this, different bilayers (made from lipids with varying tail lengths and flexibility) can be exposed to simulated early-Earth conditions: heating and cooling, dehydration, solute fluctuations. The prediction is that some bilayers will consistently survive more intervals due to the way they better align with the flow of energy around them.

If this is confirmed, it would mean membranes were not just chemically possible, they were thermodynamically favored. Selection happened not by purpose, but through physics.

The persistence model predicts that lipid bilayer configurations will be filtered based on their ability to manage environmental energy fluctuations:
\begin{itemize}
  \item Bilayers with optimal fluidity and curvature resist rupture, minimizing \( \Phi(t) \).
  \item Excessively rigid or leaky membranes dissipate energy quickly and fail to persist.
  \item Certain lipid compositions (e.g., unsaturated tail mixtures) promote persistence across temperature intervals.
  \item Thermodynamically stable bilayers will accumulate over time even in the absence of replication.
\end{itemize}

\paragraph*{Tests}
\begin{itemize}
\item\textbf{Context/Gap:} Bilayer membranes are central to protocell models, yet most analyses assume steady conditions rather than fluctuating environments.
\item\textbf{Model Action:} The thermodynamic persistence model applies the augmented inequality to bilayers under hydration and thermal cycling, with curvature, fluidity, and composition altering $y(t)$ and $\Phi(t)$.

\item\textbf{Prediction (Expectation):} Bilayers with optimal curvature and fluidity are predicted to resist rupture and minimize $\Phi(t)$, while rigid or leaky bilayers should collapse. Membranes enriched in unsaturated fatty acids should persist more reliably across intervals, leading to the accumulation of stable bilayers.

\item\textbf{Falsification:} If bilayer persistence is unaffected by curvature, fluidity, or composition - such that rigid and leaky bilayers persist as well as optimally fluid ones - the framework would be falsified in this domain.
\end{itemize}

\subsection{Interpretation of outcomes}
\begin{center}
\begin{tabular}{|p{4cm}|p{4cm}|p{4cm}|}
\hline
\textbf{Outcome pattern} & \textbf{Meaning for the model} & \textbf{Consequence} \\
\hline
\textbf{All tests affirmed} \\ 
(observations match predictions across amphiphile, surface, folding, micelle, and bilayer systems)
&
Confirms that the persistence inequality $y'(t) \ge 0$ is a valid and general descriptor of prebiotic selection.
&
The persistence inequality equation provides a unified thermodynamic basis for abiogenesis; the next step is quantitative parameterization and planetary modeling.
\\
\hline
\textbf{Some affirmed, some null}
&
Suggests the core mechanism is partly correct but system-specific. Energy-entropy coupling governs persistence only within defined regimes of diffusion and confinement.
&
Guides refinement of the $\Phi(t)$ penalty term and identification of environmental boundary conditions for prebiotic stability.
\\
\hline
\textbf{No tests affirmed}
&
Indicates that persistence cannot be explained by energy alignment alone; the inequality fails as a predictive criterion.
&
The persistence model is falsified as a general theory. Either alternative variables dominate, or selection requires kinetic heredity from the outset.
\\
\hline
\end{tabular}
\end{center}

\subsection{Scientific risk}
By stating in advance what would disprove the model, the persistence model satisfies the criterion of scientific falsifiability. The framework does not seek confirmation but exposure: if nature demonstrates that none of the defined persistence signatures occur, the theory must be abandoned or fundamentally revised. Conversely, consistent affirmation across diverse systems would strengthen the case that thermodynamic selection predates biological evolution.

While the model presented here maintains a simplified structure for clarity, it is intentionally designed to accommodate future refinements involving more complex thermodynamic behaviors.

First, the energy input function $z(t)$ may reflect environmental energy derived from various sources (e.g., photonic, thermal, or chemical). For example, a temperature-dependent function could be formulated and extended, while reaction requirements ($x_i$) and releases ($r_i$) may likewise be expressed as functions of temperature, allowing for incorporation of Arrhenius-style activation rates without disrupting the core logic of the model.

Second, stored energy is not intended to represent thermal capacity in the strict sense, but rather structural or chemical potential energy retained within the system. This distinction allows stored energy to serve as a functional buffer against environmental fluctuations, without being tied to classical specific heat or enthalpy dynamics.

Third, the model is robust under stochastic energy input. For example, if z(t) follows a Gaussian or similarly noisy distribution, stored energy acts as a compensatory term, enabling reaction chains to persist through periods of low environmental input. This introduces a selection pressure favoring systems that can store energy ahead of scarcity,  a primitive resilience mechanism that parallels biological buffering strategies.

Finally, refinements to the model explicitly incorporate entropy and spatial constraints. Local entropy reduction ($\Delta S_{local}$), environmental entropy export ($\Delta S_{env}$), and diffusion penalties ($D(x,t)$) can be used to extend the core viability inequality. These modifiers reflect thermodynamic constraints in open systems and allow for more realistic modeling of self-assembly, dissipation, and compartmentalization. The inequality thus remains general but extensible, supporting both conceptual exploration and numerical simulation.

The test case scenarios proposed here, show how the model can be experimentally tested using simple compounds and observable phenomena. They connect entropy regulation, diffusion control, and structural stability into a single framework and reinforces the core claim: persistence arises where entropy export aligns with structural coherence under fluctuating energy.

\subsection{Universal Scope and Molecular Agnosticism}

The Thermodynamic Abiogenesis Likelihood Model (TALM) is not confined to terrestrial assumptions of biology. It does not depend on carbon-based molecular structures, nucleotides, or biochemical replication. Rather, it defines the emergence of life-like systems in terms of energy viability and persistence within structured, entropy-driven environments. This positions TALM as a framework that is fundamentally molecularly agnostic and scalable across physical regimes.

By grounding the emergence of selection-like behavior in thermodynamic alignment rather than chemical specificity, TALM invites the consideration of exotic life forms, composed of silicon, metal-surface catalysis, quantum structures, or unknown substrates, as valid targets for origin-of-life investigation. The only constraint imposed is the existence of fluctuating energy, compartmentalizing structure, and the potential for local entropy reduction consistent with overall entropy production. This approach is consistent with the theoretical perspectives of Davies and Cleland, who argue that biology as currently known may only reflect a narrow subset of possible life-supporting structures \cite{cleland2002life, davies2009emergent}.

Moreover, TALM is inherently scalable. The probability function
\[
P_{\text{selective}} = 1 - (1 - P(y'(t) \geq 0))^N
\]
allows for extrapolation from localized chemical systems (\(N\) as individual reaction permutations) to whole-planetary or even galactic estimations (\(N\) as the total number of habitable micro-environments or reaction domains). This scaling supports the view that abiogenesis may be not merely possible but statistically probable across large enough thermodynamic landscapes. Similar views have been proposed in the context of cosmological selection \cite{lineweaver2002galactic, walker2016universal}, where life is treated as a common feature of entropy-utilizing complexity.

In this light, the Thermodynamic Abiogenesis Likelihood Model does not model the origin of ``life'' in a strict biological sense, but rather the origin of persistent structures where thermodynamic systems exhibit selective retention within entropic environments. It therefore provides a potential bridge between prebiotic chemistry, astrobiology, and universal principles of organized complexity.

\subsection{Implications for Abiogenesis}

In this framework, natural selection does not depend on heredity or replication at the outset, but emerges as a thermodynamic sieve: systems that maintain reaction viability under dynamic energy input $z(t)$, often aided by energy storage $y(t)$, network resilience $R_n$, and entropic compatibility, are retained, while others collapse.

This selection pressure favors not only systems that align energetically, but also those whose structure arises through entropy-driven complexity. Open systems naturally explore a vast space of molecular configurations, and entropy - through self-assembly and spontaneous ordering - acts as a creative force in generating that diversity. The energy flow then filters this space, selectively stabilizing systems that maintain coherence across energetic fluctuation.

Over time, this process preferentially retains:
\begin{itemize}
  \item Reaction networks that dissipate energy effectively \cite{england2013statistical}
  \item Autocatalytic or structurally recursive sets \cite{kauffman1993origins}
  \item Systems capable of storing energy or buffering against loss \cite{pross2012what}
  \item Structures arising from entropy-compatible self-assembly \cite{prigogine1977self, schneider2005into}
\end{itemize}

In this view, abiogenesis is not a singular event, but a gradient: a transition shaped by energy dissipation, entropic transformation, and spatial-temporal persistence. Natural selection begins not with replication, but with thermodynamic survival - an emergent property of open systems responding to energy gradients and entropic constraints.

Consequently, the Thermodynamic Abiogenesis Likelihood Model, by treating life emergence as a thermodynamic selection process, leads to several key predictive implications:
\begin{itemize}
  \item \textbf{Life-like systems are statistically inevitable} in diverse environments where fluctuating energy and chemical complexity coexist \cite{england2013statistical}.
  \item \textbf{Non-replicative prebiotic systems may be far more common} than those capable of heredity \cite{walker2013algorithmic}.
  \item \textbf{Gas giants, icy moons, and subsurface oceans} could all plausibly host persistence-based chemical systems, even in the absence of traditional biosignatures \cite{seager2013exoplanet}.
  \item \textbf{Life emergence may occur in multiple distinct epochs} on a single body, driven by cyclical thermodynamic phases rather than one-time events \cite{bains2004many}.
  \item \textbf{Exoplanetary systems will exhibit structure favoring thermodynamically stable bodies}, potentially leading to testable correlations between exoplanet class and biopotential \cite{seager2013exoplanet}.
\end{itemize}
These implications suggest a revision of existing metrics of habitability, extending consideration beyond carbon-based chemistry and stable surface liquid water

The theoretical outlook presented here positions the Thermodynamic Abiogenesis Likelihood Model as both a bridge and a filter between conceptual thermodynamics and experimental reality. By defining persistence in measurable energetic terms, the model converts the question of life's emergence from one of abstract plausibility into one of empirical exposure. Each proposed test therefore represents a deliberate opportunity for the framework to fail - a necessary criterion for genuine scientific progress. If the predicted signatures of persistence are not observed, the theory must be revised or rejected; if they are, then the principle that thermodynamic alignment precedes biological heredity gains tangible support. In this sense, tests are specifically designed to satisfy the criterion that a sound theoretical model must not only describe what might occur but also specify the precise conditions under which it would not.

\section{Conclusion}

The predictive application of the persistence inequality to aqueous amphiphilic systems demonstrates the model’s utility beyond abstract thermodynamic reasoning. By examining acetate, octanoate, and dodecanoate under identical conditions, we identify a chain-length threshold at which vesicle formation becomes thermodynamically favored. This aligns with established hydrophobic theory while reframing it through the lens of energetic persistence. It also provides a concrete and falsifiable prediction: when energy intervals and chain length surpass defined thresholds, spontaneous structural organization should appear.

Grounded in a reaction viability inequality, the framework explains how fluctuating environmental energy can drive the persistence or collapse of non-replicative molecular systems. By formalizing energy requirements ($x_i$), releases ($r_i$), storage ($s_i$), and resilience ($R_n$), it shows how differential persistence can arise in the absence of heredity or replication, offering a genuinely prebiotic mechanism of selection. In this view, selection emerges as a thermodynamic property of open systems rather than a biological innovation.

Contextualized within the foundational work of Kauffman, Prigogine, Deamer, England, and Pross on dissipative structures and self-organization, the model integrates entropy production, diffusion constraints, and spatial modulation into a physically coherent description of early prebiotic environments. The Thermodynamic Abiogenesis Likelihood Model (TALM) extends this principle to planetary scales, enabling probabilistic assessments of persistence under specific energy-cycling and spatial conditions and providing a structured way to compare the abiogenic potential of different worlds.

Crucially, TALM is scalable and agnostic to molecular identity. Its criteria apply not only to terrestrial carbon chemistry but to any structured environment where energy gradients and entropy dynamics permit the formation of persistent configurations. This broadens the plausible scope of life’s emergence across diverse planetary scenarios and challenges the assumption that abiogenesis requires rare or narrowly defined conditions.

If the predictions outlined here are verified, the origin of life is no longer an improbable chemical accident but a measurable physical outcome of energy/entropy alignment. The proposed experimental scenarios translate the abstract viability condition into operational tests, linking the persistence inequality to known systems such as micelles, folding peptides, and surface-assisted chemistry. Combined with the Thermodynamic Abiogenesis Likelihood Framework, they establish a practical, physically grounded approach for identifying prebiotically favored systems - structures selected not for biological function, but for energetic persistence.

\backmatter

\bmhead*{Supplementary information}

\bmhead*{Acknowledgments}

The author would like to express special thanks to Dr. David Deamer and Dr. Henderson 'Jim' Cleaves for their generous guidance, thoughtful critique, and encouragement to pursue experimental grounding for this theoretical framework. Their contributions helped refine both the conceptual clarity and the empirical relevance of this work.

Further thanks are extended to Professor Lee Cronin and Dr. Addy Pross for providing critical feedback on earlier drafts of the manuscript, which helped shape the direction and clarity of the final model.

\section{Declarations}

\begin{itemize}
\item Funding: Not applicable.
\item Conflict of interest: The author declares none.
\item Ethics approval and consent to participate: Not applicable.
\item Consent for publication: Not applicable.
\item Data availability: All data are included within the article and its supplementary information.
\item Materials availability: Not applicable.
\item Code availability: Not applicable.
\item Author contribution: T.M.Prosser is the sole author of this work.
\end{itemize}

\begin{appendices}

\end{appendices}

\bibliography{sn-bibliography}


\end{document}